\documentclass[times,twoside,final]{elsarticle}

\usepackage{framed,multirow}
\usepackage{latexsym}
\usepackage{url}
\usepackage{xcolor}
 \usepackage{float}
\setcitestyle{authoryear,sort&compress,open={(},close={)}}
\usepackage{hyperref}
\definecolor{newcolor}{rgb}{.8,.349,.1}
\newcommand\norm[1]{\lVert#1\rVert}

\usepackage{fancyhdr}
\fancypagestyle{plain}{%
    \fancyhead{} 
    \fancyfoot[C]{\thepage}
}
\hypersetup{pdfauthor={Name}}

\usepackage{subcaption,booktabs}
\usepackage[justification=centering]{caption}

\graphicspath{{figs/}} 
\DeclareGraphicsExtensions{.pdf,.jpeg,.jpg}
\usepackage{amsmath,amssymb,multirow}
\usepackage{mathrsfs,array,stfloats}
\UseRawInputEncoding
\usepackage{graphicx}
\begin{document}

\begin{frontmatter}

\title{Cross-dimensional transfer learning in medical image segmentation with deep learning}   

\author[1]{Hicham Messaoudi\corref{correspondingauthor}}
\ead{hicham.messaoudi@univ-bejaia.dz}
\cortext[correspondingauthor]{Corresponding author.\\}

\author[2,3]{Ahror Belaid}
\address[1]{Laboratory of Medical Informatics (LIMED), Faculty of Technology, University of Bejaia, 06000 Bejaia, Algeria}

\address[2]{Laboratory of Medical Informatics (LIMED), Faculty of Exact Sciences, University of Bejaia, 06000 Bejaia, Algeria}

\address[3]{Data Science \& Applications Research Unit - CERIST, 06000, Bejaia, Algeria}

\author[4,5]{Douraied Ben Salem}

\address[4]{Laboratory of Medical Information Processing (LaTIM) UMR 1101, Inserm, 29200, Brest, France}

\address[5]{Neuroradiology Department, University Hospital of Brest, 29200, Brest, France}

\author[4,6]{Pierre-Henri Conze}
\address[6]{IMT Atlantique, 29200, Brest, France}

\begin{abstract}
Over the last decade, convolutional neural networks have emerged and advanced the state-of-the-art in various image analysis and computer vision applications. The performance of 2D image classification networks is constantly improving and being trained on databases made of millions of natural images. Conversely, in the field of medical image analysis, the progress is also remarkable but has mainly slowed down due to the relative lack of annotated data and besides, the inherent constraints related to the acquisition process. These limitations are even more pronounced given the volumetry of medical imaging data. In this paper, we introduce an efficient way to transfer the efficiency of a 2D classification network trained on natural images to 2D, 3D uni- and multi-modal medical image segmentation applications. In this direction, we designed novel architectures based on two key principles: weight transfer by embedding a 2D pre-trained encoder into a higher dimensional U-Net, and dimensional transfer by expanding a 2D segmentation network into a higher dimension one. The proposed networks were tested on benchmarks comprising different modalities: MR, CT, and ultrasound images. Our 2D network ranked first on the CAMUS challenge dedicated to echo-cardiographic data segmentation and surpassed the state-of-the-art. Regarding 2D/3D MR and CT abdominal images from the CHAOS challenge, our approach largely outperformed the other 2D-based methods described in the challenge paper on Dice, RAVD, ASSD, and MSSD scores and ranked third on the online evaluation platform. Our 3D network applied to the BraTS 2022 competition also achieved promising results, reaching an average Dice score of 91.69\% (91.22\%) for the whole tumor, 83.23\% (84.77\%) for the tumor core and 81.75\% (83.88\%) for enhanced tumor using the approach based on weight (dimensional) transfer. Experimental and qualitative results illustrate the effectiveness of our methods for multi-dimensional medical image segmentation.
\end{abstract}


\end{frontmatter}

\section{Introduction}

Deep learning owes its exponential success to the evolution of technological equipment, learning algorithms, and the availability of large datasets. The approaches that are derived from this have evolved the state-of-the-art in several fields such as image classification~\citep{qLi2014}, image segmentation~\citep{MUrehman2020,Minaee2021}, image annotation~\citep{HCshin2016} and computer-aided diagnosis~\citep{HIsuk2013}. Currently, in computer vision, the use of convolutional neural networks (CNN) is predominant~\citep{Zegour2023}. This is mainly due to their ability to discover hierarchical representations of data, making them perfectly suitable for any image-related task. Medical imaging has been widely used over the past years for early cancer detection~\citep{Atakiddin2021,SMshah2022}, drug development~\citep{Gliang2020}, and treatment care~\citep{MJiqbal2021}. Accurate image segmentation plays a pivotal role in cancer diagnosis and can improve long-term survival rates. However, the direct interpretation of a large set of medical images can be tedious. The need for automation is more than necessary, especially with the emergence of high-dimensional data.

However, deep learning networks require a large amount of data to be able to generalize effectively. In medical imaging, there is still a latent lack of high-quality annotated data. This is mainly due to the established regulations imposed in different countries~\citep{MJwillemink2020} and the significant cost of manual annotation, the latter being even more critical when dealing with uni- or multi-modal 3D medical data. In addition, manual annotation is error-prone and time-consuming at large scale, and the development of clinically meaningful, accurate, and automatic segmentation systems is even more crucial.

One of the most promising strategies towards robust medical image segmentation deals with transfer learning~\citep{Sbozinovski2020}. Its core concept is based on the re-use of networks pre-trained on a specific task from another application in order to significantly save computational resources, accelerate convergence during training and improve the network efficiency. The underlying assumption is that many extracted features, especially low-level ones, are usually shared between different image types and tasks. Most research in 3D segmentation is currently focused on improving the training process and centered on the exploration of encoder-decoder network architectures capability~\citep{Zjiang2020,YXzhao2020}. However, very few studies~\citep{Sstarke2020,Imerino2021} have been carried out on the re-use of pre-trained 2D networks for 3D medical image segmentation, and none to our knowledge has explored the performance of cross-dimensional transfer learning. The latter designation refers to the use of a sequence of $(m+n)$-dimensional weights, extrapolated from the weights of a pre-trained $m$-dimensional network, for the initialization of a segment or a whole $(m+n)$-dimensional network.

In this paper, we propose several novel network architectures for 2D, 3D uni- and multi-modal medical image segmentation. First, we develop a 2D U-Net-like architecture using a pre-trained 2D encoder, intended for the segmentation of 2D echo-cardiographic data and 3D abdominal organs. We then present the second network of our weight transfer learning, which consists of the incorporation of a pre-trained 2D segmentation network into the core of a 3D U-Net-like architecture. In addition, we propose the dimensional transfer learning which is an extrapolation of 3D weights from 2D encoder weights. These weights are then used to initialize the encoder of a 3D U-Net-like architecture. These two networks are designed for the segmentation of 3D uni- and multi-modal data of brain tumors. In this work, the baseline 2D encoder chosen in our experiments is the EfficientNet~\citep{Mtan2019} architecture, but any other classification network could be considered. Our main contributions are summarized below\footnote[1]{\href{https://github.com/hic-messaoudi/Cross-dimensional-transfer-learning-in-medical-image-segmentation-with-deep-learning}{The code and additional resources are available at this link.}}:
\begin{itemize}
\item Proposal of a variety of ways to re-use a 2D classifier network for 2D/3D segmentation purposes.
\item The re-use principles are generic with respect to the choice of the classifier and its application.
\item Exploitation of these networks on several medical imaging modalities including ultrasound, magnetic resonance (MR), and computed tomography (CT) images.
\item Evaluation and validation of the proposed networks on various benchmarks including BraTS, CAMUS, and CHAOS, with promising results.
\end{itemize}

\subsection{Related works}

Currently, CNN architectures have largely conquered the field of image segmentation, which can be defined as the classification of a set of pixels or voxels according to a semantic criterion chosen beforehand. Thus, several architectures based on different approaches have been designed, including the encoder-decoder architecture which remains dominant and constitutes the basis of the majority of the currently proposed frameworks. In medical image segmentation, the aim is to automatically or semi-automatically delineate healthy anatomical or pathological tissues for a variety of purposes ranging from simple assisted diagnosis to image-guided surgery.

A well-known network architecture and the most popular one for these applications is the U-Net~\citep{Oronneberger2015} architecture. This network is relatively fast and mitigates the limitations due to data scarcity issues. Its architecture is based on an encoder, which is a contracting path used for the compression and interpretation of the data into an internal latent representation. The latter is linked to a symmetrical decoder which is applied to regain spatial coherence. Motivated by scene understanding applications, SegNet~\citep{Vbadrinarayanan2017} appeared after and is made of an encoder architecture that is topologically similar to VGG-16 and uses skip-connections as in U-Net. A major distinction between them is that SegNet uses less memory since skip-connections from the encoder derive from the compressed feature maps by pooling layers whereas U-Net uses full-scale feature maps. The attention U-Net~\citep{Ooktay2018att} introduces a novel attention gate (AG) model, motivated by the empirical observation of its usefulness in extracting relevant features and its ability to automatically learn to focus on target structures of varying morphology. U-Net++~\citep{Zzhou2020} is an architecture that aims at improving the vanilla U-Net and takes mainly advantage of redesigned skip-connections, where the feature maps of the encoder are enriched by intermediate convolution layers. A deep supervision scheme was also used for better optimization of the learning process of the decoder and to get an earlier representation of output predictions. As an improvement to U-Net++, U-Net3+ was developed ~\citep{Hhuang2020} based on full-scale skip-connections to explore sufficient information and better estimate both organ position and morphology. The architecture was designed with reduced parameter complexity compared to U-Net++ to improve the computational efficiency.

Furthermore, during the different segmentation-based competitions, many architectures have been developed to tackle a variety of challenges. In the brain tumor segmentation competition (BraTS) 2020~\citep{Sbakas2017,Sbakas2018,BHmenze2015} and to enable diversity in prediction results, ~\cite{Thenry2021} presented two networks with similar architecture inspired by the 3D U-Net~\citep{Ocicek2016} using two different training approaches. The initial convolutional block was set to 48 channels for better use of the original spatial information before compression. They used two sets of networks along with a deep supervision scheme and a stochastic weight averaging~\citep{Pizmailov2018} approach to improve the predictive power of their networks. Another 3D U-Net-based architecture was developed by ~\cite{Pahmad2021} with reduced parametric complexity and residual Inception blocks to learn multi-scale contexts. Dense connections were also used with a constrained growth-rate to limit unnecessary features. Dilated convolutions were employed in the encoder to expand the size of the receptive field in the feature maps leading to promising results. ~\cite{Yyuan2021} introduced SA-Net, a 3D U-Net inspired architecture that replaces the long-range skip connections between the same scale with full-scale skip-connections to make full use of original feature maps. SA-Net came with a dynamic scale attention mechanism that ensures a correct appreciation of the significance of each feature map.

However, the previously described networks are usually randomly trained, which can be sub-optimal in terms of computation time and performance, and for tasks involving a limited amount of annotated data. The weight initialization scheme used being a critical aspect for the generalization of deep neural networks, several works have shown a way to overcome those problems through the use of pre-trained network weights arising from larger datasets. The fact that these encoders were not trained on a closer domain to the targeted task does not prevent the clear improvement of training and test scores. The superiority of this approach has been empirically confirmed when compared to the use of random initialization. With TernausNet, a classical U-Net along with incorporated pre-trained VGG-11, ~\cite{IVladimir1} showed a significant improvement in network performance compared to using random initialization on aerial imaging data~\citep{aerodata}. In the second version of TernausNet, the authors replaced the encoder with a pre-trained ABN WideResnet-38~\citep{BSamuel} and showed superior results~\citep{IVladimir2} in comparison to other methods employed on DeepGlobe-CVPR~\citep{Deepglob}, a challenge dedicated to the segmentation, classification, and detection of satellite images.~\cite{conze2020healthy} extended this idea to the medical field by using a VGG-16 pre-trained on ImageNet~\citep{imagenet} as part of the encoder of a U-Net architecture (v16U-Net). The results showed a substantial gain in performance over classical U-Net and fully randomly initialized v16U-Net for pathological shoulder muscle MR segmentation purposes. The same findings arose with a deeper network based on VGG-19 in \citep{conze2021abdominal} for abdominal multi-organ (liver, kidneys, spleen) segmentation from CT scans.

\begin{figure*}[!ht]
\begin{minipage}[]{\hsize}
\centering
\includegraphics[width=\linewidth]{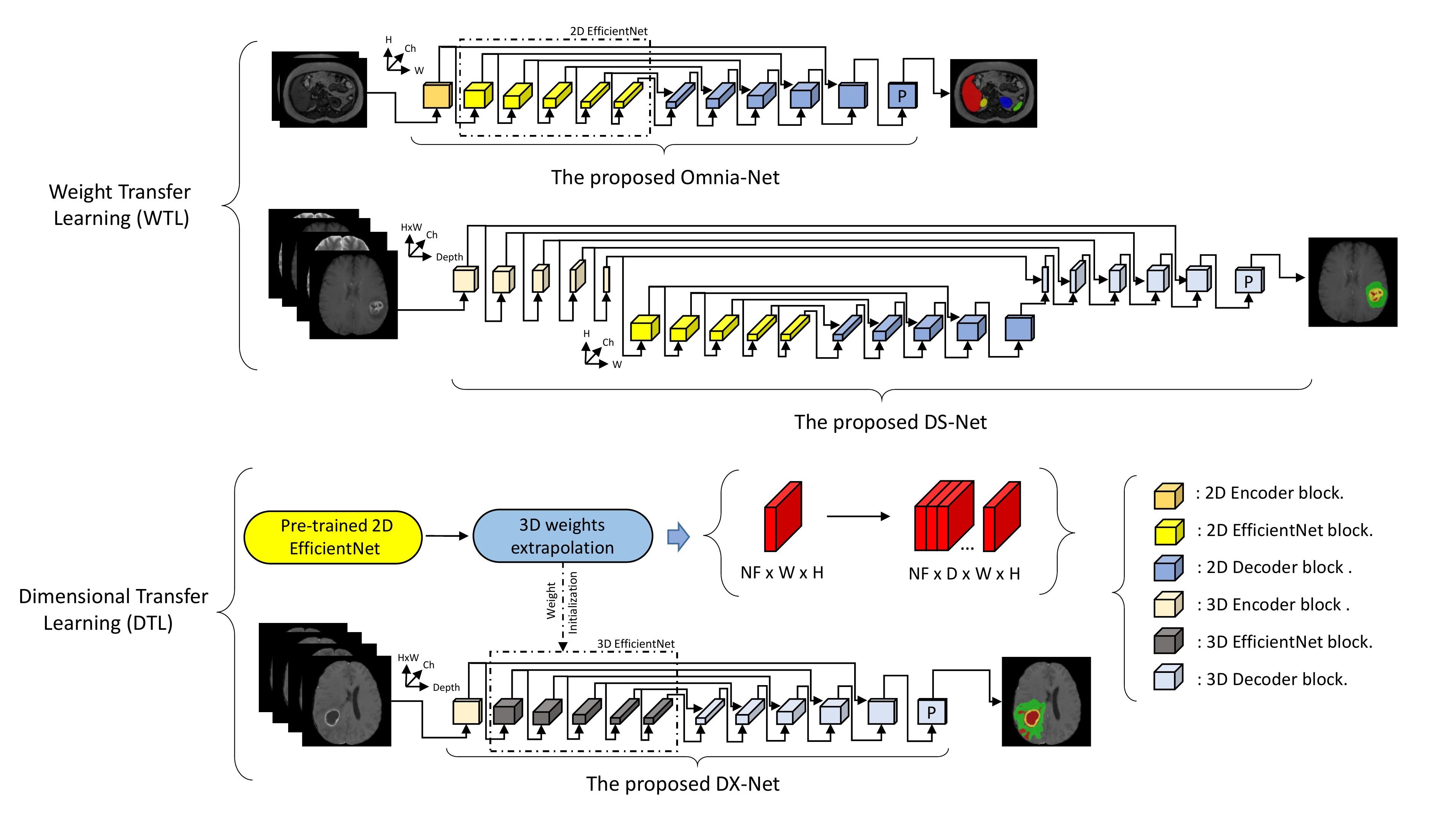}
\end{minipage} \vspace{-0.4cm}
\caption{Graphical summary of the proposed approaches and architectures. From top to bottom : weight transfer learning (WTL) approach with the proposed 2D (Omnia-Net) and 3D (DS-Net) segmentation networks. Below is the dimensional transfer learning (DTL) approach along with an instance of its application using the proposed DX-Net.}
\label{fig:summary_methods}
\end{figure*}

One of the most promising strategies in network modeling developed in recent years deals with the introduction of a scaling method that uniformly adjusts depth, width, and resolution dimensions by using a compound coefficient. Therefore, a new family of architectures referred to as EfficietNet~\citep{Mtan2019} was created through neural architecture search~\citep{Bzoph2016} and demonstrated outstanding accuracy and parameter efficiency on image classification tasks. These networks are also highly effective when transferred to an image segmentation context and several works have been carried out on this path.~\cite{Qyuan2021} proposed EfficientU-Net++ for the segmentation of melanoma skin lesions, which is a combination of a U-Net++ and a pre-trained EfficientNet with re-designed skip-connections for the aggregation of features from varying scales. EfficientUNet++ showed higher scores than U-Net or U-Net++ in that particular task.~\cite{LDhuynh2020} used the same strategy for the detection and segmentation of artifacts and diseases in endoscopy images and replaced the U-Net++ encoder with an EfficientNet towards more accurate feature extraction. They also highlighted the influence of test-time augmentation (TTA) on the improvement of segmentation results. 

~\cite{Jwang2021} introduced EAR-U-Net for automatic liver segmentation in CT, which is a 2D U-Net like architecture using EfficientNet-B4 for the encoder part and an attention gate scheme in the skip-connections to discriminate the useful features from the irrelevant ones. With the aim of developing a fully-automated and efficient COVID-19 detection system, ~\cite{NKchowdhury2021} proposed the ECOVNet architecture, which uses an EfficientNet with weights pre-trained on ImageNet. Its classification performance was promising and showed a clear improvement using different ensemble strategies along with a visualization technique that is provided to highlight the data features related to class distinction. In our previous work~\citep{Hmessaoudi2021}, by including the EfficientNet architecture as part of the encoding branch, we proposed an asymmetric 3D U-Net architecture for the segmentation of brain tumors. The first layers of the encoder were devoted to reducing the depth dimension to fit the 2D EfficientNet input. Experimental results on validation and test data showed that the proposed method achieves promising performance.

However, the weights produced by EfficientNet in a supervised training context can be improved, as shown by~\cite{Qxie2020} through noisy student training, which is a semi-supervised learning technique that, unlike knowledge distillation~\citep{Aromero2015,Ghinton2015}, uses a student network of a same or larger size with added noise. In order to take advantage of the weights of existing pre-trained 2D networks, ~\cite{Imerino2021} proposed to derive their 3D architectural versions for classification purposes. The comparison between 2D and 3D classification networks, with and without transferred weights showed a significant improvement when extrapolating 3D weights from 2D, up to a score difference ratio of 18\% for the 3D version of the Inception ResNet architecture.

\begin{table}[t]
\scriptsize
\centering
\begin{tabular}{lcccc}
\hline
{Method} &
{Network} &
{\begin{tabular}[c]{@{}c@{}}Domain\end{tabular}} &
{Dataset} &
{Ranking} \\ \hline
\begin{tabular}[l]{@{}l@{}}Weight Transfer \\ Learning\end{tabular} &
\begin{tabular}[c]{@{}c@{}}\\Omnia-Net\\ \\DS-Net\end{tabular} &
\begin{tabular}[c]{@{}c@{}}2D echo-cardiography  \\ 3D abdominal \\ \\ 3D multi-modal \\brain tumors\end{tabular} &
\begin{tabular}[c]{@{}c@{}}CAMUS \\  CHAOS\\ \\ BraTS\end{tabular} &
\begin{tabular}[c]{@{}c@{}}1$^{st}$\\ 3$^{rd}$\\ \\ -\end{tabular} \\ \hline
\begin{tabular}[L]{@{}l@{}}Dimensional Transfer \\ Learning\end{tabular} &
\begin{tabular}[c]
{@{}c@{}}DX-Net \end{tabular} &
\begin{tabular}[c]
{@{}c@{}}3D multi-modal \\brain tumors 
\end{tabular} &
BraTS &
- \\ \hline
\end{tabular}
\caption{Summary of the different proposed approaches and networks and their ranking according to the dataset used and their domain of application.} \vspace{-0.2cm}
\label{tab:meth}
\end{table}

The limitations of the above-mentioned methods for image segmentation are that they either require a large amount of data to train a network from scratch or that they are limited in terms of performance using the 2D weights transfer. In this work, we propose a novel method for 2D and 3D medical image segmentation that overcomes these limitations. In Sect.~\ref{sec:material}, we describe the proposed approaches followed by the architectures derived from them. We start with the weight transfer learning, where we introduce the Omnia-Net architecture for 2D echo-cardiographic segmentation and 3D segmentation of abdominal organs (CT, MR), along with the DS-Net architecture for brain tumor segmentation in a 3D multi-modal setting. In the same context, we introduce the dimensional transfer learning using the DX-Net architecture. In Sect.~\ref{sec:results}, we display the results obtained through statistical and graphical evidences to show the effectiveness of our approaches. We discuss the findings and report our interpretations before concluding in Sect.~\ref{sec:conclusion}.

\section{Material and methods}
\label{sec:material}

We introduce in this section the two proposed transfer learning principles that are: weight and dimensional transfer learning, from which three networks are derived. The first one aims at re-using the weights of a 2D classification network for 2D medical image segmentation tasks and can be extended to adjacent problems. The second network is defined as the integration of a 2D segmentation architecture into a 3D one and is intended for 3D brain tumor segmentation. The last one can be defined as the 3D transformation of 2D weights from a classification network, and their use for the initialization of the encoder of a 3D U-Net-like network  (Tab.~\ref{tab:meth}).

\subsection{Weight transfer learning}

In the following section, we introduce the weight transfer learning (WTL) approach, which can be formulated as a generalization of standard transfer learning. The WTL approach is based on the re-use of a pre-trained network on a different task or paradigm without any alteration of its weights in the process. This can be performed through classical transfer learning such as the re-use of a 2D network on a different classification task, or its incorporation into a network designed to address a different problem (as in the case for segmentation networks with pre-trained encoders), or even its incorporation into a higher dimensional network

\begin{figure*}[!ht]
\begin{minipage}[]{\hsize}
\centering
\includegraphics[width=\linewidth]{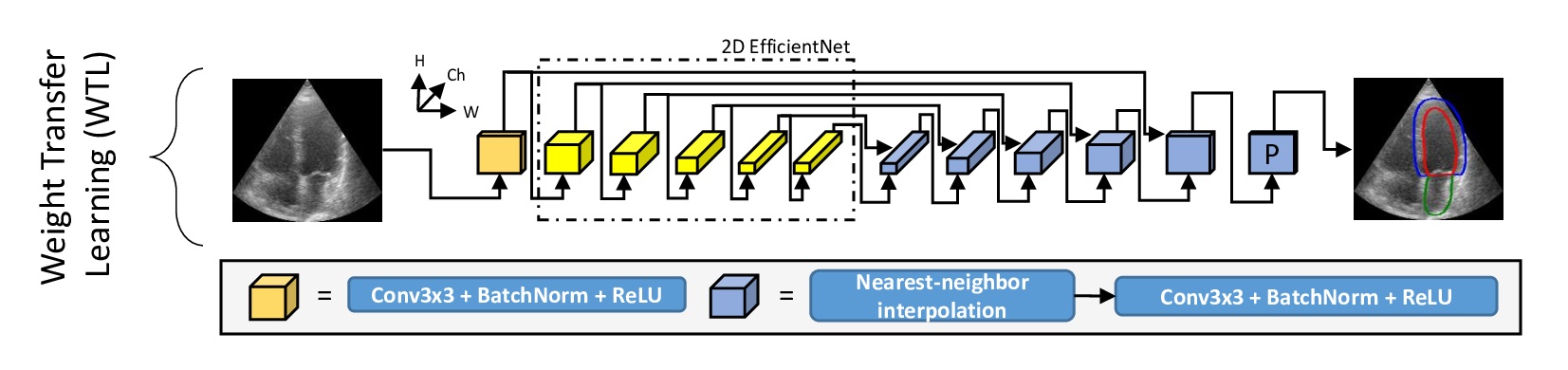}
\end{minipage} 
\caption{Schematic illustration of the proposed 2D network architecture (Omnia-Net).}
\label{fig:omnianet}
\end{figure*}

\begin{figure*}[!ht]
\begin{minipage}[]{\hsize}
\centering
\includegraphics[width=\linewidth]{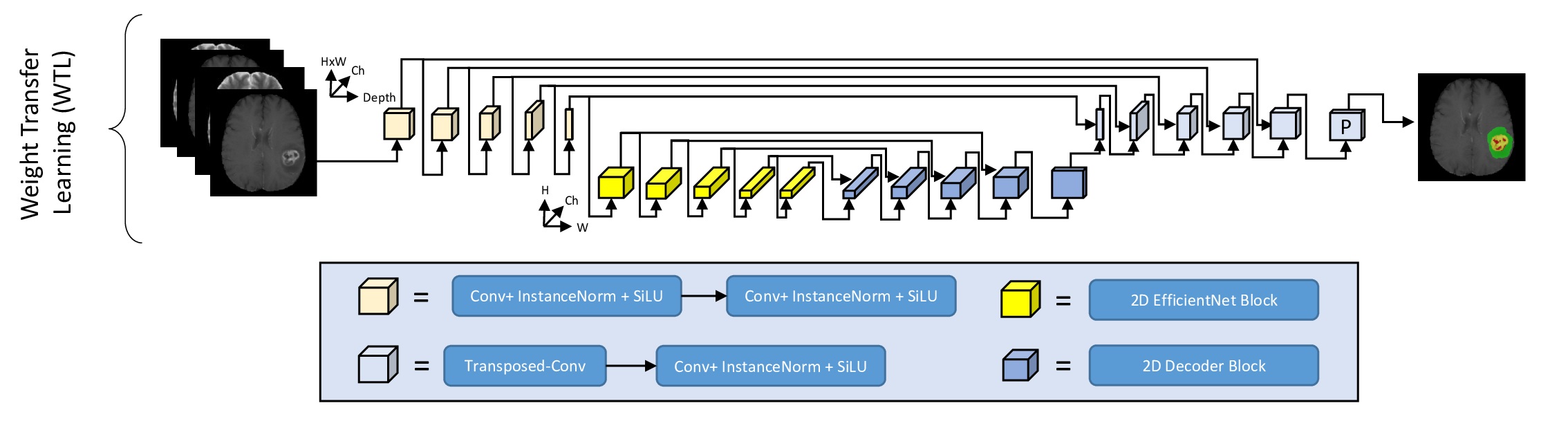}
\end{minipage} \vspace{-0.4cm}
\caption{Schematic illustration of the proposed DS-Net architecture. The intra-slice encoder as well as the decoder consists of a succession of convolutional blocks with SiLU-activated instance normalization.}
\label{fig:ds_net}
\end{figure*}

We introduce the use of a pre-trained 2D classification network, as an encoder of a U-Net-like architecture, which we refer to as Omnia-Net due to its suitability for use in both 2D and 3D contexts and its maintained performance in various modalities. The Omnia-Net encoder architecture is slightly modified to incorporate a convolutional block to take advantage of the full-scale characteristics of the input images as shown in Fig.~\ref{fig:omnianet}. We evaluate its efficiency and validate its potential for the segmentation of 2D echo-cardiographic data from the CAMUS dataset~\citep{Sleclerc2019} and in the context of 3D abdominal organ MR and CT segmentation using the CHAOS dataset~\citep{AEkavur2019,AEkavur2021}. Note that only the encoder part of Omnia-Net is pre-trained. Finally, we incorporate Omnia-Net into a higher dimensional architecture, the dimensionally-stacked network (DS-Net) as shown in Fig.~\ref{fig:ds_net}. The DS-Net is intended for 3D multi-modal brain tumor segmentation. We assess its performance using the BraTS 2022 dataset~\citep{brats2022}.

\subsubsection{2D echo-cardiographic image segmentation}\label{sssec:num1}

To further support the empirical evidence of the generalizability power of pre-trained networks over randomly initialized ones, we validate our network using several datasets. CAMUS is the first used for this purpose. It is the largest publicly available fully-annotated 2D echo-cardiographic dataset. It consists of clinical images acquired from 500 different patients with different chamber views using optimized acquisition techniques to enable the evaluation of left ventricular ejection fraction measurements along with a wide variability of acquisition settings. 
\newcommand{\mysize}{3.5}
\begin{figure}[!h]
\begin{minipage}[h]{\hsize}
\centering

\includegraphics[height=\mysize cm,width=\mysize cm]{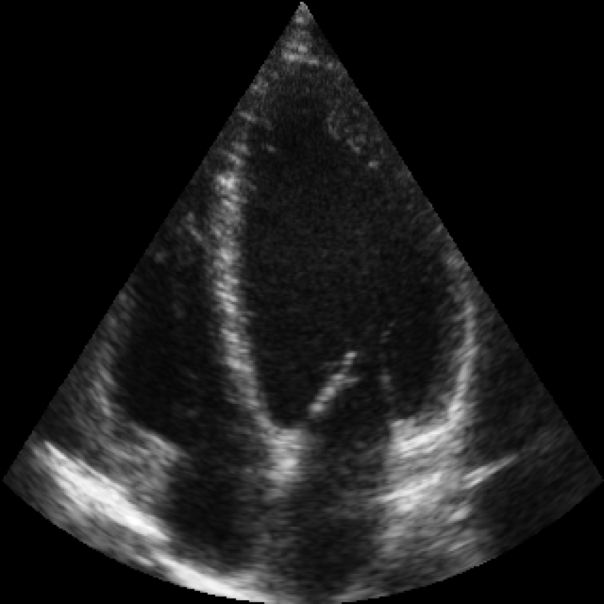}
\includegraphics[height=\mysize cm,width=\mysize cm]{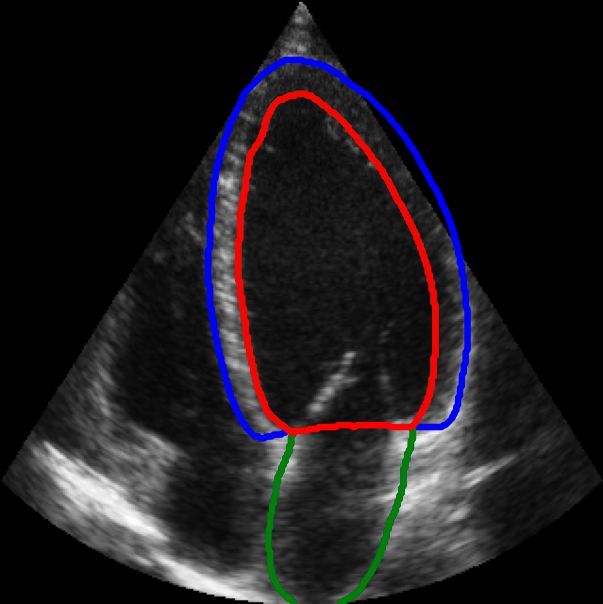}
\vspace{1mm}
\end{minipage}
\begin{minipage}[]{\hsize}
\centering
\includegraphics[height=\mysize cm, width=\mysize cm]{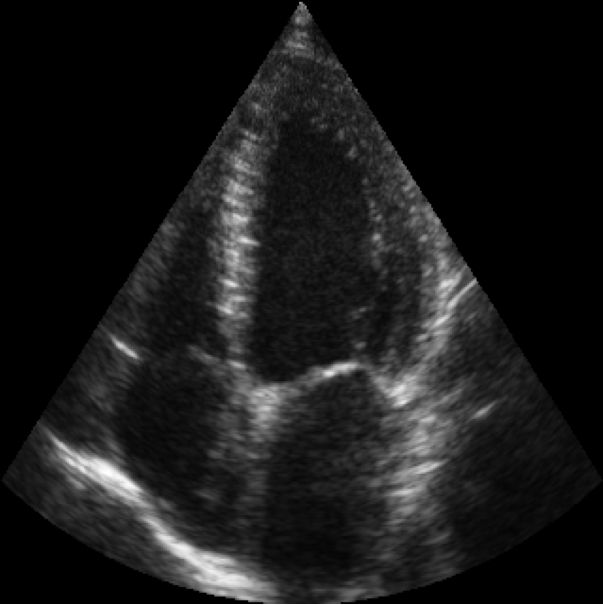}
\includegraphics[height=\mysize cm, width=\mysize cm]{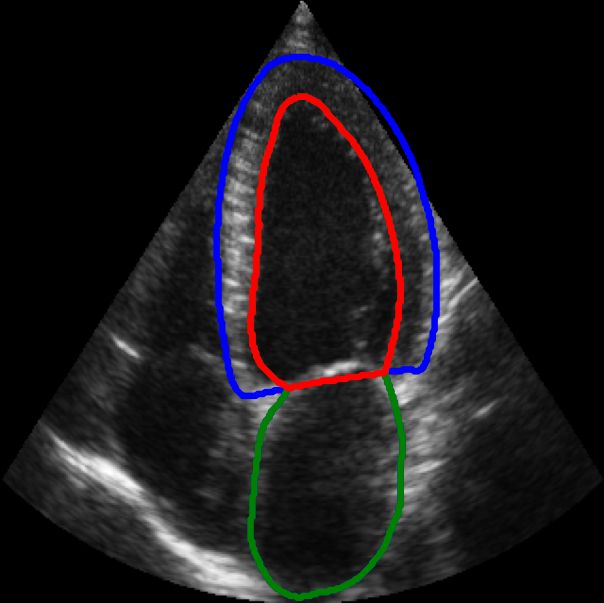}
\vspace{1mm}
\end{minipage}
\caption{Visualizations of ED (top) and ES (down) samples of the CAMUS test set. From left to right: echo-cardiographic data and its corresponding predicted annotation by Omnia-Net. The epicardium and endocardium of the left ventricle and left atrium wall are shown respectively in blue, red, and green.}
	\label{fig:camustest}
\end{figure}

The dataset is divided into images containing end diastole (ED) and end systole (ES) phases for both 2 and 4 chambers views, each data including information relative to the characteristics and quality of the images. Three cardiologists ($O_{1}$, $O_{2}$, $O_{3}$) manually annotated the test set, while $O_{1}$ manually annotated the entire dataset (train and test). Also, the challenge organizers did not stop at the representation of inter-observer variability but were also interested in the intra-observer variability by providing manual annotation of $O_{1}$'s test set at 7-month interval ($O_{1a}$, $O_{1b}$). Samples of the test set can be seen in Fig.~\ref{fig:camustest}.

For deep learning methods, U-Net 1 and U-Net 2~\citep{Sleclerc2019} were trained by the challenge creators. Regarding these two variants of the original U-Net, one is optimized for speed with 2M parameters and the other one for accuracy with 17.5M parameters. U-Net 1 is used as a segmentation module in the anatomically constrained neural network (ACNN)~\citep{Ooktay2018acnn}. The stacked Hourglass network~\citep{Anewell2016} has also been used for comparison purposes, it included a set of 3 encoder-decoder networks along with a deep supervision scheme. The final segmentation results were defined by the output of the third network. The architecture of the U-Net++ has been adapted to obtain the best results on this particular dataset. The same data pre-processing and post-processing methods have been applied to all the mentioned networks~\citep{Sleclerc2019}, contrary to our methodology where no post-processing is employed.

In our work, we focus on using the EfficientNet-B0 as a baseline. More complex versions of the network could also be used, taking into account the adaptation of both resolution and regularization (dropout, more aggressive data augmentation) schemes. We opt for the implementation of~\citep{pim} for the encoder part. Each layer of the decoder is a succession of nearest neighbor interpolation, concatenation, two convolutional layers with a kernel size of 3, each followed by a batch normalization and ReLU activation. For data pre-processing, we opt for an independent normalization of each sample of the dataset from which we subtract the mean of the image, divided by the standard deviation. The output activation function of the network is a Softmax in our case.

\subsubsection{3D abdominal organ MR segmentation}

\begin{figure}[!h]
\centering
\begin{minipage}[]{\hsize}
\centering
\includegraphics[height=\mysize cm, width=\mysize cm]{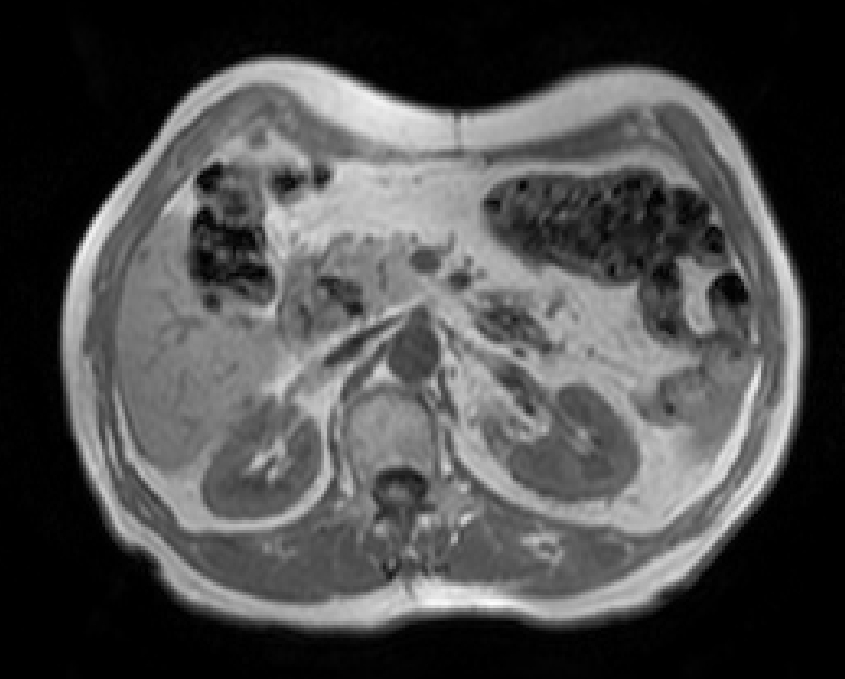}
\includegraphics[height=\mysize cm, width=\mysize cm]{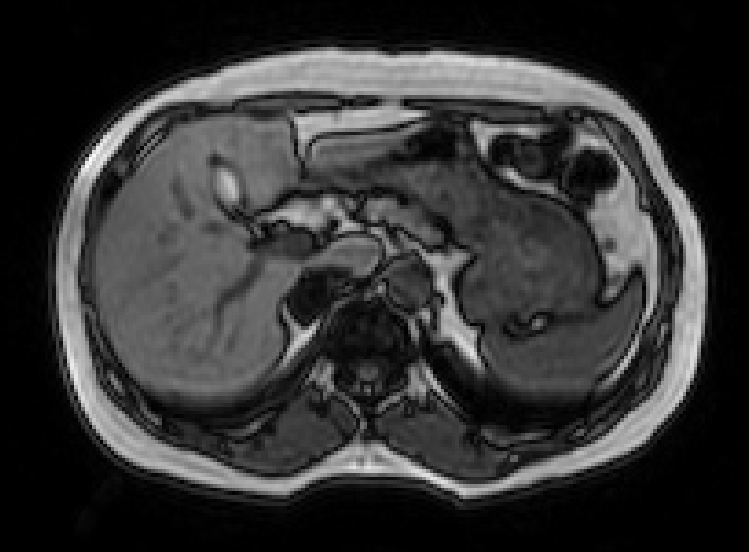}
\vspace{1mm}
\end{minipage}
\begin{minipage}[]{\hsize}
\centering
\includegraphics[height=\mysize cm, width=\mysize cm]{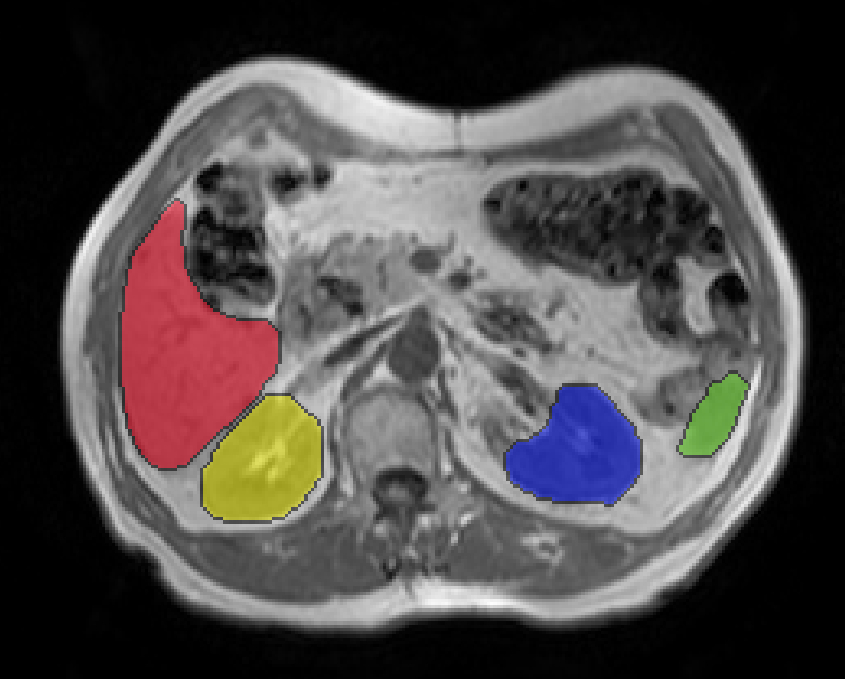}
\includegraphics[height=\mysize cm, width=\mysize cm]{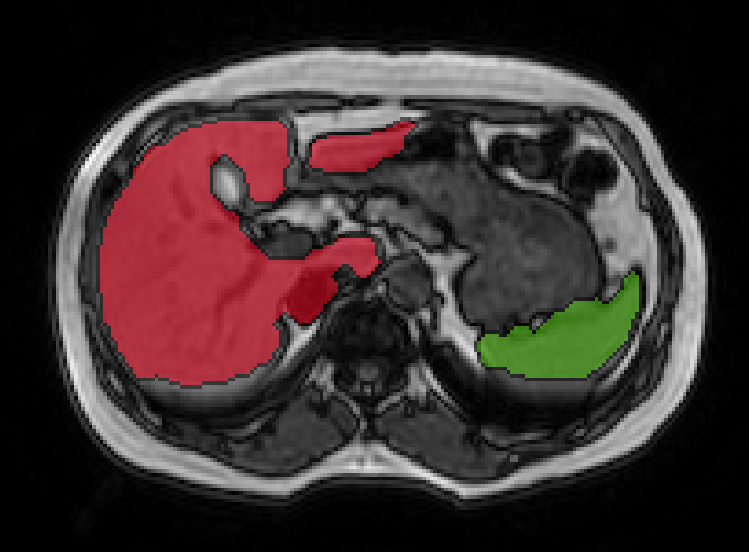}
\vspace{1mm}
\end{minipage}
\caption{Visualizations on two samples from the CHAOS training set~\citep{AEkavur2021}. From left to right: liver, right and left kidney, and spleen in MR T1-in and T1-out modalities and their corresponding delineations are shown respectively in red, yellow, blue, and green.}
\label{fig:Chaos1}
\end{figure}

During the last years, several deep learning methods have been proposed for the segmentation of abdominal organs. The most common approaches are based on U-Net \citep{Oronneberger2015}, fully convolutional networks \citep{Shelhamer2017} or Mask R-CNN \citep{Jin2021}. These methods show a very good performance in terms of segmentation quality but require large amounts of annotated data to achieve this level of performance. Recently, the CHAOS dataset~\citep{AEkavur2021} has been created to assess the performance of automatic organ segmentation algorithms on abdominal CT and MR images. It contains annotated abdominal organ data from 80 healthy patients, composed of 40 CT and MR data consisting of two different sequences (T1 and T2). Two samples from the test set can be seen in Fig.~\ref{fig:Chaos1}. The CHAOS challenge is divided into 5 tasks. Task 1 is a multi-modal binary segmentation of the liver, the goal being to create a system able to take as inputs CT or MR images of different sequences and to infer their correct masks using a single system. Tasks 2 and 3 are binary CT and MR liver segmentation tasks, respectively. Task 5 represents a generalization of task 3 to other organs (left and right kidneys, spleen). Task 4 is a generalization of task 1 to other organs. In this context, the CT images include only the liver class, while the MR images include the four abdominal organs. 

Commonly, using a 2D network to segment higher dimensional data may seem counter-intuitive due mainly to the loss of spatial coherence. Nevertheless,~\cite{Fisensee2017} showed a clear performance degradation using a 3D network for the segmentation of anisotropic data. In addition, several 2D architectures during the CHAOS competition were employed, as is the case for~\cite{conze2021abdominal} using a conditional generative adversarial network and Ernst et al., with a modified Attention 2D U-Net~\citep{ANabila} along with multi-scaled input image pyramid to enhance feature representation. Pham et al used a different approach with an architecture composed of three modules (auto-encoder, Hourglass network, 2D U-Net), this scheme being used to enhance the organ localization capabilities of the automated approach. These methods have all performed more or less efficiently, some better than others on the different challenge tasks. This shows the effectiveness of these approaches in specific instances but does not really show the particular element that made them perform better on certain tasks when compared to others.

\begin{figure*}[!ht]
\begin{minipage}[]{\hsize}
\centering
\includegraphics[width=\linewidth]{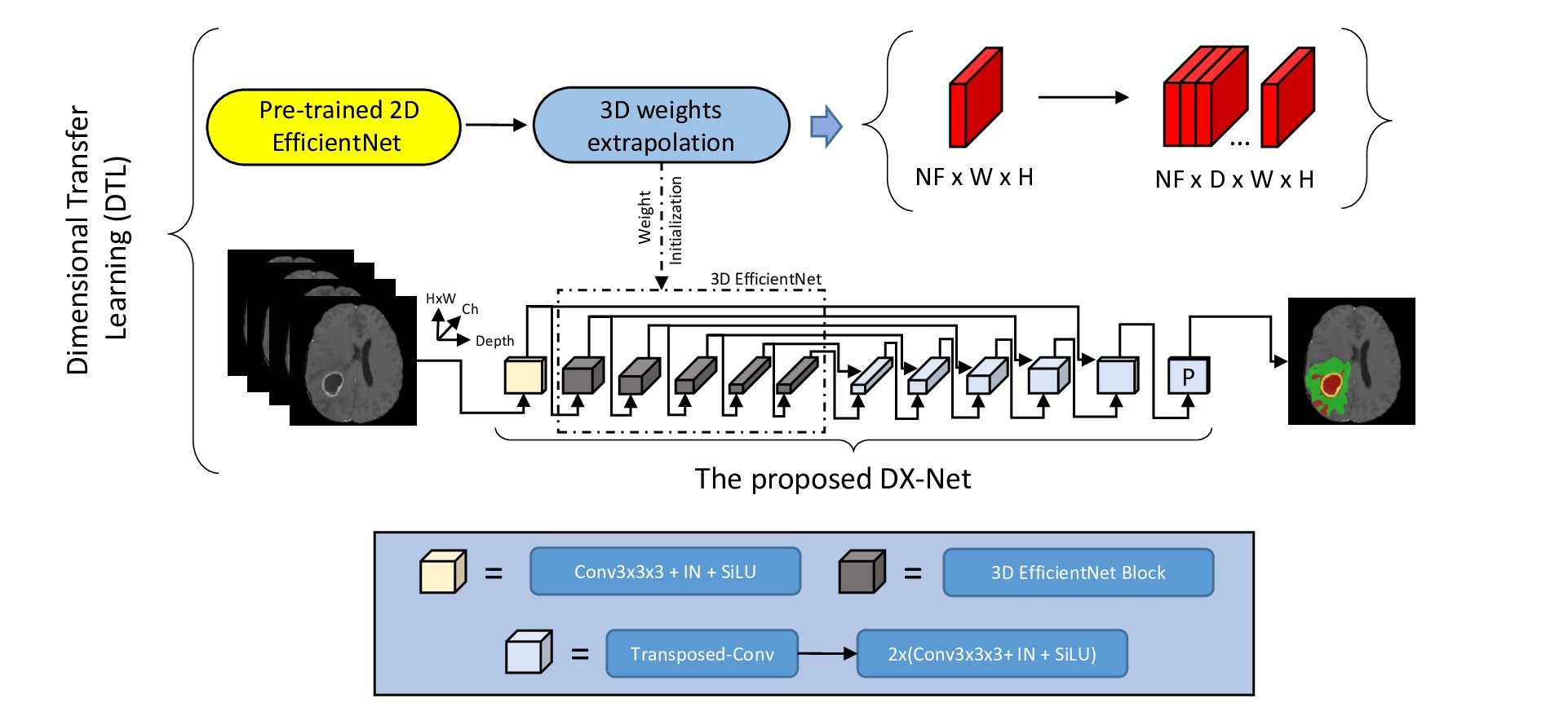}
\end{minipage}\\
\caption{Schematic illustration of the proposed DX-Net architecture. T-Conv and IN respectively stand for transposed convolution~\citep{Eshelhamer2016} and instance normalization~\citep{Dulyanov2016}.}
\label{fig:effunet3d}
\end{figure*}

\begin{figure}[!h]
\begin{minipage}[]{\hsize}
\centering
\includegraphics[height=\mysize cm, width=\mysize cm]{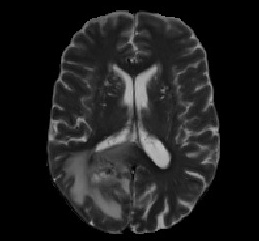}
\includegraphics[height=\mysize cm, width=\mysize cm]{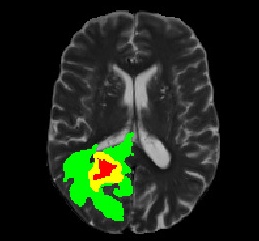}
\vspace{1mm}
\end{minipage}
\begin{minipage}[]{\hsize}
\centering
\includegraphics[height=\mysize cm, width=\mysize cm]{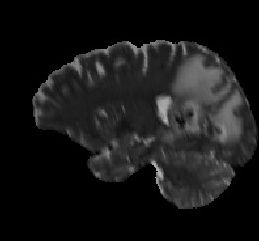}
\includegraphics[height=\mysize cm, width=\mysize cm]{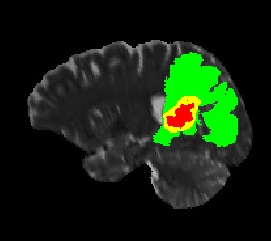}
\vspace{1mm}
\end{minipage}
\begin{minipage}[]{\hsize}
\centering
\includegraphics[height=\mysize cm, width=\mysize cm]{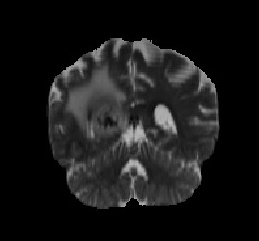}
\includegraphics[height=\mysize cm, width=\mysize cm]{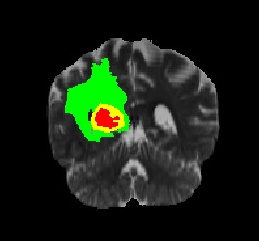}
\vspace{1mm}
\end{minipage}
\caption{Visualizations on the $440^\textrm{th}$ MR T2 data of the BraTS 2022 training set~\citep{brats2022}. From top to bottom: axial, sagittal, and coronal views are shown with their corresponding ground truth. Edema is shown in green, enhancing tumor in yellow, and necrosis/non-enhancing tumor in red.}
\label{fig:Brats2}
\end{figure}

\subsubsection{3D multi-modal brain tumor MR segmentation}

Brain tumor segmentation is a crucial task for monitoring patient diagnosis and correct disease isolation. In order to establish a concrete context for the listing and ranking of the various algorithms and approaches proposed for this purpose, the multi-modal brain tumor segmentation (BraTS) 2022 dataset~\citep{brats2022} was created. The dataset includes a large number of low- and high-grade gliomas data obtained from routine multi-institutional multi-modal MR imaging scans, subdivided into training, validation, and test sets. Each data volume consists of 4 multi-modal scans which are native (T1), post-contrast T1-weighted (T1ce), T2-weighted (T2), and T2 Fluid Attenuated Inversion Recovery (FLAIR) volumes. The different regions considered for the evaluation include: the enhancing tumor (ET) which is depicted by the areas that show hyper-intensity in T1ce when compared to T1, the tumor core (TC) which is the part that is typically resected surgically and includes the ET region as well as the necrotic (fluid-filled) and non-enhancing (solid) regions of the tumor core (NET/Ncr), the latter being hypo-intense in T1ce when compared to T1 and the whole tumor (WT) which is the combined TC region and the peritumoral edema (ED), that can be identified by the hyper-intense signal in FLAIR. Training samples can be seen in Fig.~\ref{fig:Brats2}. 

As shown in Fig.~\ref{fig:ds_net}, two phases comprise the encoding and decoding processes. First, we encode 3D data into 2D data, preserving the original dimensions of height and width and just compressing the depth to a limited number of channels. Second, the data is being processed to be adapted as an input to the 2D EfficientNet, which is the second encoding stage. The output of the 2D pre-trained encoder is a tensor with a batch size of 4 and 3 channels which is transmitted to a 2D decoder consisting at each level of an upsampling layer processing a nearest neighbor interpolation followed by a skip-connection and two blocks comprising a convolutional layer, a batch normalization layer, and a ReLU activation function. The output of the 2D decoder is not activated and is transmitted after adequate transformation to the 3D decoder. The latter follows the same structure as the 2D decoder, except that it uses instance normalization layers instead of batch normalization, which is considered much more adequate given the used batch size for the original data. 

\subsection{Dimensional transfer learning}

In this section, instead of processing data as 2D axial images, we directly employ volumetric medical images including multiple channels in a multi-modal image segmentation context. We propose a novel image segmentation architecture based on dimensional transfer learning referred to as Dimensionally-eXpanded network (DX-Net). As displayed in Fig.~\ref{fig:effunet3d}, DX-Net is a U-Net-like architecture including a modified 3D EfficientNet encoder. The initialization of its weights is done using the weights of its 2D equivalent. Each set of weights is translated into 3D by concatenating the 2D weights repeatedly over the 3D depth. The 2D weights are drawn from a noisy-student training~\citep{Qxie2020} on natural images (ImageNet). The particular choice of these weights is motivated by our belief that a network with solid performance on natural image classification will be better suited to perform feature extraction as an encoder for an image segmentation task. Therefore, we hypothesize a performance transfer across adjacent tasks. The DX-Net decoder, on the other hand, is randomly initialized and is described in much more detail in the following section. DX-Net is intended for 3D multi-modal medical image segmentation. We evaluate its performance using the 3D MR brain tumor segmentation dataset (BraTS).

\subsubsection{3D multi-modal brain tumor MR segmentation}

Motivated by the favorable results obtained by the weight transfer of pre-trained 2D encoders in the context of the segmentation of higher dimensional medical images, we propose in this part a straightforward approach to re-use the weights of those networks. It consists of the transformation of pre-trained classification network 2D weights into 3D. The produced weights are used to initialize the equivalent 3D architecture parameters. This process has already been applied in the context of classification~\citep{Imerino2021} and yields excellent results when compared to a random initialization, the difference being that our objective is to use this dimensional transfer approach in an image segmentation context, and also apply the weights resulting from a noisy student training, instead of those acquired through a classical training on ImageNet. 

The field of 3D medical data segmentation already suffers from a latent lack of annotated data and costly computational power. Therefore, this approach represents a favorable opportunity to drastically reduce the cost and training time of higher dimensional convolutional networks. 
In the network depicted in Fig.~\ref{fig:effunet3d}, the first block is a convolution layer followed by an instance normalization and a Sigmoid linear unit (SiLU) activation function. The main idea is to capture the spatial features of the inputs before down-sampling in order to improve the reconstruction of the segmented regions through skip-connections. The remaining part of the encoder is a 3D version of EfficientNet-B0, the choice of the architecture being mainly motivated by the low resolution of MR data. For the initialization of this part of the network, we use the noisy student weights of its 2D counterpart. The transformation of these weights into 3D is performed by projecting the information of the 2D learnable parameters onto the depth of their 3D counterpart. The batch normalization layers are all replaced by instance normalization layers since we are using a batch size of 1. Each block of the decoder comprises a transposed 3D convolution layer followed by a succession of two blocks consisting of 3$\times$3 convolution layer, instance normalization, and SiLU activation. The output layer is followed by a Sigmoid activation to binarize the predictions. The training strategy is the same as the one used in the weight transfer section.

\section{Results and discussion}
\label{sec:results}

In this section, we introduce the results of the methods mentioned in Sect.~\ref{sec:material} by following the same order of presentation.

\subsection{Weight transfer learning}

We start by reporting the results obtained with the weight transfer learning approach on 2D echo-cardiographic and 3D abdominal data as well as 3D multi-modal brain tumor data, while discussing the obtained results.

\subsubsection{2D echo-cardiographic image segmentation}
\label{sssec:echo}

In the original paper~\citep{Sleclerc2019}, CAMUS is presented as a variable image quality dataset. In our work, we show experimentally that data quality has little to no impact on the performance of model optimization as long as the data is uniformly annotated.

\begin{figure}[!h]
  \begin{minipage}[]{\hsize}
		\centering
		\textbf{Manual annotation :}  Training set - Patient 009 ED 2Ch.\\
		\vspace{1mm}
		\includegraphics[height=\mysize cm, width=\mysize cm]{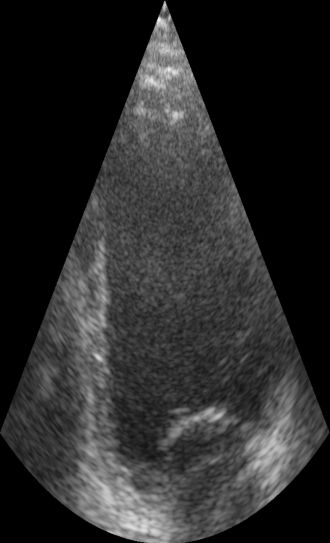}
		\includegraphics[height=\mysize cm, width=\mysize cm]{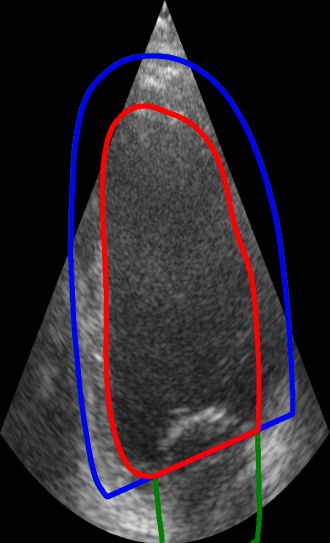}
		\vspace{1mm}
	\end{minipage}
	\begin{minipage}[]{\hsize}
		\centering
		\textbf{Best :} Test set - Patient 029 ED 4Ch - Dice Scores : Endo : 0.97. Epi : 0.98. LA : 0.93.\\
		\vspace{1mm}
		\includegraphics[height=\mysize cm, width=\mysize cm]{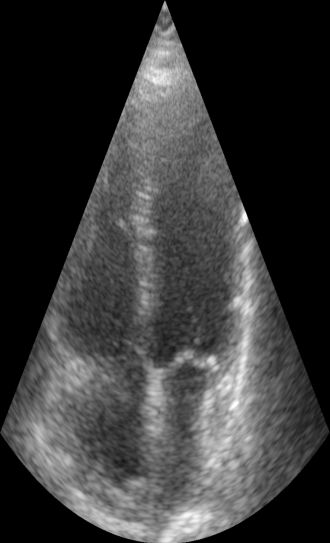}
		\includegraphics[height=\mysize cm, width=\mysize cm]{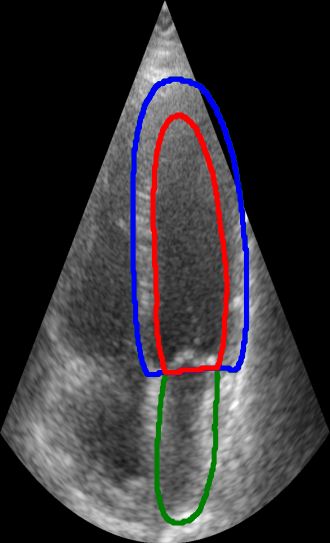}
		\vspace{1mm}
	\end{minipage}
	\begin{minipage}[]{\hsize}
		\centering
		\textbf{Worse :} Test set - Patient 050 ED 2Ch - Dice Scores : Endo : 0.92. Epi : 0.92. LA : 0.68.\\
		\vspace{1mm}
		\includegraphics[height=\mysize cm, width=\mysize cm]{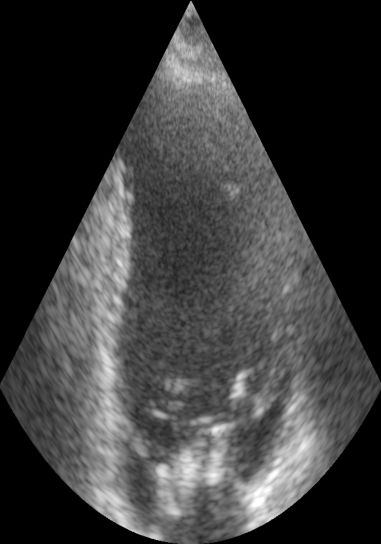}
		\includegraphics[height=\mysize cm, width=\mysize cm]{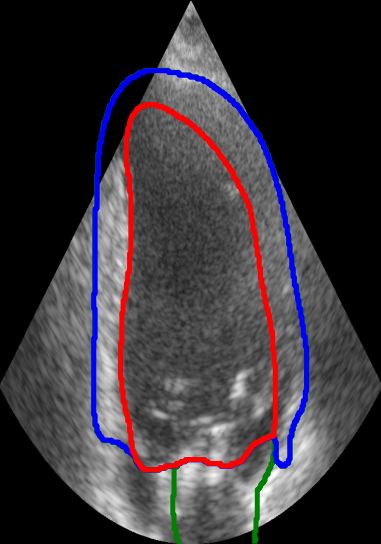}
		\vspace{1mm}
	\end{minipage}
	\caption{Qualitative results on CAMUS dataset~\citep{Sleclerc2019}. Cases were selected as manual annotations from the training set, best and worst from the test set. Within each row: input data 2- or 4-chambers view at end-diastolic (ED) phase and their corresponding annotations on the right where epicardium and endocardium of the left ventricle and left atrium wall are shown respectively in blue, red, and green.}
	\label{fig:camus_last_fig}
\end{figure}

Several metrics are used to evaluate the performance of the proposed methods, on image segmentation and volume estimation of multi-chamber view ultrasound images. Dice score is a standard metric for evaluating the performance of image segmentation methods. The Dice score is defined as follows:

\begin{equation}
    Dice(A,B) = \frac{2|A\cap B|}{|A|+|B|},
\end{equation}

\noindent where $|.|$ denotes the cardinality, $A$ the segmentation mask of the ground truth and $B$ the predicted segmentation mask. The Dice score ranges between 0 and 1, with a higher value indicating better performance. In addition, the mean absolute distance (MAD) is used for performance evaluation. For $a_{i}\in A$ and $b_{i} \in B$, MAD$(A,B)$ is given by: 

\begin{equation}
    \text{MAD}(A,B) = \frac{1}{N}\sum_{i=1}^{N}|a_i-b_i|,
\end{equation}

\noindent where $N$ refers to the total number of samples, and  $|.|$ the absolute value. Lower MAD values indicate a more accurate segmentation. The Hausdorff distance (HD)  is also used to evaluate the performance of segmentation methods. HD is defined as:

\begin{equation}
    HD(A,B) = \max(h(A,B), h(B,A))
\end{equation}

\noindent where : 

\begin{equation}
    h(A,B) = \max_{a\in A}\min_{b\in B}\norm{a-b}
\end{equation}

\noindent A smaller HD value indicates a better similarity between predicted segmentation results and ground truth.

\begin{table*}[!ht]
\centering
\resizebox{\linewidth}{!}{%
\begin{tabular}{ccccccccccccc}
\hline
\multirow{3}{*}{{Methods}} & \multicolumn{6}{c}{{ED}}                                       & \multicolumn{6}{c}{{ES}}                  \\ \cline{2-13} 
 &
  \multicolumn{3}{c}{{$\textrm{LV}_\textrm{Endo}$}} &
  \multicolumn{3}{c}{{$\textrm{LV}_\textrm{Epi}$}} &
  \multicolumn{3}{c}{{$\textrm{LV}_\textrm{Endo}$}} &
  \multicolumn{3}{c}{{$\textrm{LV}_\textrm{Epi}$}} \\ \cline{2-13} 
 &
{Dice} &
{MAD} &
{HD} &
{Dice} &
{MAD} &
{HD} &
{Dice} &
{MAD} &
{HD} &
{Dice} &
{MAD} &
{HD} \\ \hline
$O_{1a}$ vs $O_{2}$               & 0.919 & 2.2 & 6.0  & 0.913 & 3.5          & \multicolumn{1}{c|}{8.0}  & 0.873 & 2.7          & 6.6  & 0.890 & 3.9 & 8.6  \\
$O_{1a}$ vs $O_{3}$               & 0.886 & 3.3 & 8.2  & 0.943 & 2.3          & \multicolumn{1}{c|}{6.5}  & 0.823 & 4.0          & 8.8  & 0.931 & 2.4 & 6.4  \\
$O_{2}$ vs $O_{3}$                & 0.921 & 2.3 & 6.3  & 0.922 & 3.0          & \multicolumn{1}{c|}{7.4}  & 0.888 & 2.6          & 6.9  & 0.885 & 3.9 & 8.4  \\
$O_{1a}$ vs $O_{1b}$              & 0.945 & 1.4 & 4.6  & 0.957 & 1.7          & \multicolumn{1}{c|}{5.0}  & 0.930 & 1.3          & 4.5  & 0.951 & 1.7 & 5.0  \\ \hline
SRF                               & 0.895 & 2.8 & 11.2 & 0.914 & 3.2          & \multicolumn{1}{c|}{13.0} & 0.848 & 3.6          & 11.6 & 0.901 & 3.5 & 13.0 \\
BEASM-fully                       & 0.879 & 3.3 & 9.2  & 0.895 & 3.9          & \multicolumn{1}{c|}{10.6} & 0.826 & 3.8          & 9.9  & 0.880 & 4.2 & 11.2 \\
BEASM-semi                        & 0.920 & 2.2 & 6.0  & 0.917 & 3.2          & \multicolumn{1}{c|}{8.2}  & 0.861 & 3.1          & 7.7  & 0.900 & 3.5 & 9.2  \\ \hline
U-Net 1                           & 0.934 & 1.7 & 5.5  & 0.951 & 1.9          & \multicolumn{1}{c|}{5.9}  & 0.905 & 1.8          & 5.7  & 0.943 & 2.0 & 6.1  \\
U-Net 2                           & 0.939 & 1.6 & 5.3  & 0.954 & \textbf{1.7} & \multicolumn{1}{c|}{6.0}  & 0.916 & \textbf{1.6} & 5.5  & 0.945 & 1.9 & 6.1  \\
ACNN                              & 0.932 & 1.7 & 5.8  & 0.950 & 1.9          & \multicolumn{1}{c|}{6.4}  & 0.903 & 1.9          & 6.0  & 0.942 & 2.0 & 6.3  \\
SHN                               & 0.934 & 1.7 & 5.6  & 0.951 & 1.9          & \multicolumn{1}{c|}{5.7}  & 0.906 & 1.8          & 5.8  & 0.944 & 2.0 & 6.0  \\
U-Net++                          & 0.927 & 1.8 & 6.5  & 0.945 & 2.1          & \multicolumn{1}{c|}{7.2}  & 0.904 & 1.8          & 6.3  & 0.939 & 2.1 & 7.1  \\
Omnia-Net (ours) &
  \textbf{0.945} &
  \textbf{1.5} &
  \textbf{4.8} &
  \textbf{0.956} &
  1.8 &
  \multicolumn{1}{c|}{\textbf{5.4}} &
  \textbf{0.922} &
  \textbf{1.6} &
  \textbf{4.8} &
  \textbf{0.951} &
  \textbf{1.8} &
  \textbf{5.9} \\ \hline
\end{tabular}}
\caption{Comparison of image segmentation scores of non-deep and deep learning methods from the CAMUS challenge and the proposed Omnia-Net architecture on CAMUS test set~\citep{Sleclerc2019}. $\textrm{LV}_\textrm{En}$: Endocardial contour of the left ventricle; $\textrm{LV}_\textrm{Ep}$: Epicardial contour of the left ventricle; ED: End diastole; ES: End systole; Dice : Dice Score; MAD: Mean Absolute Distance; HD: Hausdorff Distance.
The values in bold refer to the best performance for each measure.}
\label{tab:camusdice}
\end{table*}

During training, we reserve 80\% of the dataset while randomly allocating 20\% for the validation set and we also employ a 5-fold cross-validation scheme to perform an optimal training such that the network can take advantage of all existing characteristics in the training set. The Nadam optimizer is employed, which is none other than Adam with Nesterov accelerated gradient. This choice is motivated by the fact that it is theoretically and in most cases empirically superior to Adam~\citep{Sruder2017}. We opt for an initial learning rate of 3 $\times$ 10$^{-4}$ which is reduced at each epoch by 5\% until it reaches 1 $\times$ 10$^{-5}$, the latter being the critical learning rate defining the end of the training. We chose to use a compound loss function~\citep{Sjadon2020} combining Dice and binary cross-entropy (BCE), which is adequate for tasks that require dealing with unbalanced classes (see Eq.~\ref{eq:eq1}).

\begin{equation} \label{eq:eq1}
	\begin{aligned} 
		&\mathscr{L}_{loss} =  1 - \frac{2}{N} \sum_{n=0}^{N} \frac{\sum_{i=0}^{I} y_{i,n} g_{i,n}}{\sum_{i=0}^{I}{y_{i,n}} + \sum_{i=0}^{I}{g_{i,n}} + \epsilon} - \\ 
		& \frac{1}{N} \sum_{n = 0}^{N}{\sum_{i=0}^{I} {y_{i,n}~\log{(g_{i,n})} + (1 - y_{i,n})~\log{(1-g_{i,n})}}}\enspace,
	\end{aligned}
\end{equation}

\noindent where $N$ is the number of classes, $I$ the total number of spatial coordinates, $y_{i,n}$ is the Sigmoid activated output of the network and $g_{i,n}$ the binary ground truth with respect to class $n$ and $i^{th}$ data coordinate. $\epsilon$ is a small constant to avoid divisions by zero.

Moreover, we do not use any additional data for network training as we restrict our study to the provided one and we do not perform any data augmentation due to material and time constraints. To measure the reliability of the network during the training, we reserve a random validation set defined as 20\% of the dataset being thus large enough to ensure the correct estimation of the model generalization.

\begin{table}[!h]
	\begin{subtable}[]{\linewidth}
	\centering
		\begin{tabular}{cccc}
		\hline
		\multirow{2}{*}{\textbf{Methods}} & \multicolumn{3}{c}{\textbf{Corr. coefficient}}                                                                 \\ \cline{2-4} 
																			& \textbf{$\textrm{LV}_\textrm{EF}$} & \textbf{$\textrm{LV}_\textrm{EDV}$} & \textbf{$\textrm{LV}_\textrm{ESV}$} \\ \hline
		$O_{1a}$ vs $O_{2}$  & 0.801 & 0.940 & 0.956 \\
		$O_{1a}$ vs $O_{3}$  & 0.646 & 0.895 & 0.860 \\
		$O_{2}$ vs $O_{3}$   & 0.569 & 0.926 & 0.916 \\
		$O_{1a}$ vs $O_{1b}$ & 0.896 & 0.978 & 0.981 \\ \hline
		SRF                  & 0.465 & 0.755 & 0.827 \\
		BEASM-fully          & 0.731 & 0.704 & 0.713 \\
		BEASM-semi           & 0.790 & 0.886 & 0.880 \\ \hline
		U-Net 1              & 0.791 & 0.947 & 0.955 \\
		U-Net 2              & 0.823 & 0.954 & 0.964 \\
		ACNN                 & 0.799 & 0.945 & 0.947 \\
		SHN                  & 0.770 & 0.943 & 0.938 \\
		U-Net++             & 0.789 & 0.946 & 0.952 \\ \hline
		Omnia-Net (ours)                              & \textbf{0.896}                     & \textbf{0.980}                      & \textbf{0.974}                      \\ \hline
		\end{tabular}%
	\end{subtable}
	\caption{Comparison of volume estimation scores of non-deep and deep learning methods from the CAMUS challenge and the proposed Omnia-Net architecture on CAMUS test set~\cite{Sleclerc2019}. The values in bold refer to the best performance for each measure.}
	\label{tab:camusvolume}
\end{table}

For comparison purposes, we draw directly on the scores established in the original challenge. The non-deep learning methods that obtained the best results during the CETUS challenge~\citep{Obernard2016} are also included~\citep{Jpedrosa2017}. The results that can be seen in Tab.~\ref{tab:camusdice} and Tab.~\ref{tab:camusvolume} are obtained from the online evaluation platform\footnote{\href{http://camus.creatis.insa-lyon.fr/challenge/}{{http://camus.creatis.insa-lyon.fr/challenge/}}} on the test set, for which the ground truth is not available to the general public. As can be seen in Tab.~\ref{tab:camusdice}, our network clearly outperforms state-of-the-art methods for endocardial (+0.6\% for Dice, +0.1 for MAD, +0.5 for HD) and epicardial (+0.2\% for Dice, +0.3 for HD) left ventricle contour delineations in the end diastole phase. This improvement is also noticeable in the end systole phase for endocardial (+0.6\% for Dice, +0.7 for HD) and epicardial (+0.6\% for Dice, +0.1 for MAD, +0.1 for HD) left ventricle contour delineations. 

In Tab.~\ref{tab:camusvolume}, the correlation coefficient shows ostentatiously the gain in performance introduced by Omnia-Net for the ejection fraction (+7.3\%), end-diastolic (+2.6\%), and end-systolic (+1\%) ventricular volume estimation. The closeness of Omnia-Net scores to those of inter- and intra-observer contours and volumes estimation is also pertinent and shows that the method can already be valuable for clinical use. We acknowledge that the training process can be further enhanced and even expanded, given the fact that the network is trained on less than 50 epochs. Nevertheless, it proves the generalization capability of our network with a limited number of epochs and its ability to accurately reproduce manual annotations with high fidelity, as can be seen in Fig.~\ref{fig:camus_last_fig}.

\subsubsection{3D abdominal organ MR and CT segmentation}

In this section, we evaluate the proposed method on the five tasks of the CHAOS challenge. Network training is done using the 2D axial slices of the CT and MR volumes. We use the same image size throughout the challenge tasks (512 $\times$ 512). We employ several data augmentations including scaling, rotation, translation, shearing, window width/level, and additive Gaussian noise with a probability of occurrence of 0.5 for each. The B4 version of the EfficientNet is preferred, thus making a compromise between efficiency and computational complexity. It is also more appropriate for the selected image size. Regarding the parameters optimization, we exploit the same compound loss function described in Sect.~\ref{sssec:echo}.

\begin{table*}[!t]
\resizebox{\textwidth}{!}{%
\begin{tabular}{ccccccccccc}
\hline
Scores &
  Team Name &
  \textbf{Mean Score} &
  \textbf{DICE} &
  \textbf{DICE Score} &
  \textbf{RAVD (\%)} &
  \textbf{RAVD Score} &
  \textbf{ASSD (mm)} &
  \textbf{ASSD Score} &
  \textbf{MSSD (mm)} &
  \textbf{MSSD Score} \\ \hline
\multirow{6}{*}{Task 1} &
  Limed (ours) &
  \textbf{73.62 $\pm$ 12.66} &
  \textbf{0.96 $\pm$ 0.02} &
  \textbf{96.08 $\pm$ 1.82} &
  \textbf{2.78 $\pm$ 2.28} &
  \textbf{50.10 $\pm$ 33.21} &
  \textbf{1.45 $\pm$ 1.61} &
  \textbf{90.36 $\pm$ 10.76} &
  \textbf{27.50 $\pm$ 23.28} &
  \textbf{57.92 $\pm$ 25.43} \\
 &
  OvGUMEMoRIAL &
  55.78 $\pm$ 19.20 &
  0.88 $\pm$ 0.15 &
  83.14 $\pm$ 28.16 &
  13.84 $\pm$ 30.26 &
  24.67 $\pm$ 31.15 &
  11.86 $\pm$ 65.73 &
  76.31 $\pm$ 21.13 &
  57.45 $\pm$ 67.52 &
  31.29 $\pm$ 26.01 \\
 &
  PKDIA &
  50.66 $\pm$ 23.95 &
  0.85 $\pm$ 0.26 &
  84.15 $\pm$ 28.45 &
  6.65 $\pm$ 6.83 &
  21.66 $\pm$ 30.35 &
  9.77 $\pm$ 23.94 &
  75.84 $\pm$ 28.76 &
  46.56 $\pm$ 45.02 &
  42.28 $\pm$ 27.05 \\

 &
  IITKGP-KLIV &
  40.34 $\pm$ 20.25 &
  0.72 $\pm$ 0.31 &
  60.64 $\pm$ 44.95 &
  9.87 $\pm$ 16.27 &
  24.38 $\pm$ 32.20 &
  11.85 $\pm$ 16.87 &
  50.48 $\pm$ 37.71 &
  95.43 $\pm$ 53.17 &
  7.22 $\pm$ 18.68 \\ \hline
\multirow{5}{*}{Task 2} &
  Limed (ours) &
  \textbf{82.98 $\pm$ 6.12} &
  0.98 $\pm$ 0.00 &
  \textbf{98.08 $\pm$ 0.3} &
  1.48 $\pm$ 0.95 &
  70.31 $\pm$ 19.0 &
  \textbf{0.67 $\pm$ 0.12} &
  \textbf{95.54 $\pm$ 0.79} &
  \textbf{19.2 $\pm$ 12.69} &
  \textbf{68.0 $\pm$ 21.14} \\
 &
  PKDIA &
  82.46 $\pm$ 8.47 &
  0.98 $\pm$ 0.00 &
  97.79 $\pm$ 0.43 &
  \textbf{1.32 $\pm$ 1.302} &
  \textbf{73.6 $\pm$ 26.44} &
  0.89 $\pm$ 0.36 &
  94.06 $\pm$ 2.37 &
  21.89 $\pm$ 13.94 &
  64.38 $\pm$ 20.17 \\
 &
  OvGUMEMoRIAL &
  61.13 $\pm$ 19.72 &
  0.90 $\pm$ 0.21 &
  90.18 $\pm$ 21.25 &
  9 x $10^3$ $\pm$ 4 x $10^3$ &
  44.35 $\pm$ 35.63 &
  4.89 $\pm$ 12.05 &
  81.03 $\pm$ 20.46 &
  55.99 $\pm$ 38.47 &
  28.96 $\pm$ 26.73 \\

 &
  IITKGP-KLIV &
  55.35 $\pm$ 17.58 &
  0.92 $\pm$ 0.22 &
  91.51 $\pm$ 21.54 &
  8.36 $\pm$ 21.62 &
  30.41 $\pm$ 27.12 &
  27.55 $\pm$ 114.04 &
  81.97 $\pm$ 21.88 &
  102.37 $\pm$ 110.9 &
  17.50 $\pm$ 21.79 \\ \hline
\multirow{6}{*}{Task 3} &
  Limed (ours) &
  \textbf{72.88 $\pm$ 12.15} &
  \textbf{0.95 $\pm$ 0.02} &
  \textbf{95.29 $\pm$ 1.55} &
  \textbf{3.12 $\pm$ 2.38} &
  \textbf{46.16 $\pm$ 31.92} &
  \textbf{1.42 $\pm$ 0.96} &
  \textbf{90.55 $\pm$ 6.37} &
  \textbf{25.37 $\pm$ 17.17} &
  \textbf{59.51 $\pm$ 23.54} \\
 &
  PKDIA &
  70.71 $\pm$ 6.40 &
  0.94 $\pm$ 0.01 &
  94.47 $\pm$ 1.38 &
  3.53 $\pm$ 2.14 &
  41.8 $\pm$ 24.85 &
  1.56 $\pm$ 0.68 &
  89.58 $\pm$ 4.54 &
  26.06 $\pm$ 8.20 &
  56.99 $\pm$ 12.73 \\

 &
  OvGUMEMoRIAL &
  41.15 $\pm$ 21.61 &
  0.81 $\pm$ 0.15 &
  64.94 $\pm$ 37.25 &
  49.89 $\pm$ 71.57 &
  10.12 $\pm$ 14.66 &
  5.78 $\pm$ 4.59 &
  64.54 $\pm$ 24.43 &
  54.47 $\pm$ 24.16 &
  25.01 $\pm$ 20.13 \\
 &
  IITKGP-KLIV &
  34.69 $\pm$ 8.49 &
  0.63 $\pm$ 0.07 &
  46.45 $\pm$ 1.44 &
  6.09 $\pm$ 6.05 &
  43.89 $\pm$ 27.02 &
  13.11 $\pm$ 3.65 &
  40.66 $\pm$ 9.35 &
  85.24 $\pm$ 23.37 &
  7.77 $\pm$ 12.81 \\ \hline
\multirow{5}{*}{Task 4} &
  Limed (ours) &
  \textbf{61.80  $\pm$ 16.12} &
  \textbf{0.90 $\pm$ 0.06} &
  85.45 $\pm$ 20.83 &
  \textbf{7.6 $\pm$ 5.48} &
  \textbf{25.29  $\pm$ 24.71} &
  \textbf{2.98 $\pm$ 2.81} &
  80.63 $\pm$ 18.24 &
  \textbf{29.71 $\pm$ 21.22} &
  55.81 $\pm$ 26.5 \\
 &
  PKDIA &
  49.63 $\pm$ 23.25 &
  0.88 $\pm$ 0.21 &
  \textbf{85.46 $\pm$ 25.52} &
  8.43 $\pm$ 7.77 &
  18.97 $\pm$ 29.67 &
  6.37 $\pm$ 18.96 &
  \textbf{82.09 $\pm$ 23.96} &
  33.17 $\pm$ 38.93 &
  \textbf{56.64 $\pm$ 29.11} \\
 &
  OvGUMEMoRIAL &
  43.15 $\pm$ 13.88 &
  0.85 $\pm$ 0.16 &
  79.10 $\pm$ 29.51 &
  5 x $10^3$ $\pm$ 5 x $10^4$ &
  12.07 $\pm$ 23.83 &
  5.22 $\pm$ 12.43 &
  73.00 $\pm$ 21.83 &
  74.09 $\pm$ 52.44 &
  22.16 $\pm$ 26.82 \\
 &
  IITKGP-KLIV &
  35.33 $\pm$ 17.79 &
  0.63 $\pm$ 0.36 &
  50.14 $\pm$ 46.58 &
  13.51 $\pm$ 20.33 &
  15.17 $\pm$ 27.32 &
  16.69 $\pm$ 19.87 &
  40.46 $\pm$ 38.26 &
  130.3 $\pm$ 67.59 &
  8.39 $\pm$ 22.29 \\ \hline
\multirow{6}{*}{Task 5} &
  Limed (ours) &
  \textbf{69.32 $\pm$ 8.54} &
  0.93 $\pm$ 0.03 &
  91.76 $\pm$ 6.82 &
  7.67 $\pm$ 4.97 &
  22.38 $\pm$ 19.49 &
  1.46 $\pm$ 1.5 &
  \textbf{91.19 $\pm$ 6.45} &
  \textbf{18.83 $\pm$ 11.11} &
  \textbf{71.94 $\pm$ 12.14} \\
 &
  PKDIA &
  66.46 $\pm$ 5.81 &
  \textbf{0.93 $\pm$ 0.02} &
  \textbf{92.97 $\pm$ 1.78} &
  \textbf{6.91 $\pm$ 3.27} &
  \textbf{28.65 $\pm$ 18.05} &
  \textbf{1.43 $\pm$ 0.59} &
  90.44 $\pm$ 3.96 &
  20.1 $\pm$ 5.90 &
  66.71 $\pm$ 9.38 \\
 &
  OvGUMEMoRIAL &
  44.34 $\pm$ 14.92 &
  0.79 $\pm$ 0.15 &
  64.37 $\pm$ 32.19 &
  76.64 $\pm$ 122.44 &
  9.45 $\pm$ 11.98 &
  4.56 $\pm$ 3.15 &
  71.11 $\pm$ 18.22 &
  42.93 $\pm$ 17.86 &
  39.48 $\pm$ 16.67 \\
 &
  IITKGP-KLIV &
  25.63 $\pm$ 5.64 &
  0.56 $\pm$ 0.06 &
  41.91 $\pm$ 11.16 &
  13.38 $\pm$ 11.2 &
  11.74 $\pm$ 11.08 &
  18.7 $\pm$ 6.11 &
  35.92 $\pm$ 8.71 &
  114.51 $\pm$ 45.63 &
  11.65 $\pm$ 13.00 \\ \hline
\multirow{5}{*}{Average} &
  Limed (ours) &
  \textbf{72.12 $\pm$ 11.12} &
  \textbf{0.94 $\pm$ 0.03} &
  \textbf{93.33 $\pm$ 6.26} &
  \textbf{4.53 $\pm$ 3.21} &
  \textbf{42.85 $\pm$ 25.67} &
  \textbf{1.6 $\pm$ 1.4} &
  \textbf{89.65 $\pm$ 8.52} &
  \textbf{24.12 $\pm$ 17.09} &
  \textbf{62.63 $\pm$ 21.75} \\
 &
  PKDIA &
  63.98 $\pm$ 13.58 &
  0.92 $\pm$ 0.10 &
  90.97 $\pm$ 11.51 &
  5.37 $\pm$ 4.26 &
  36.94 $\pm$ 25.87 &
  4.0 $\pm$ 8.9 &
  86.40 $\pm$ 12.72 &
  29.56 $\pm$ 22.4 &
  57.4 $\pm$ 19.69 \\
 &
  OvGUMEMoRIAL &
  49.11 $\pm$ 17.87 &
  0.85 $\pm$ 0.16 &
  76.35 $\pm$ 29.67 &
  2.8 x $10^3$ $\pm$ 1 x $ 10^4$ &
  20.13 $\pm$ 23.45 &
  6.46 $\pm$ 19.59 &
  73.20 $\pm$ 21.21 &
  56.99 $\pm$ 40.09 &
  29.38 $\pm$ 23.27 \\
 &
  IITKGP-KLIV &
  38.27 $\pm$ 13.95 &
  0.69 $\pm$ 0.20 &
  58.13 $\pm$ 25.13 &
  10.24 $\pm$ 15.09 &
  25.12 $\pm$ 24.95 &
  17.6 $\pm$ 32.11 &
  49.90 $\pm$ 23.18 &
  105.57 $\pm$ 60.13 &
  10.51 $\pm$ 17.71 \\ \hline
\end{tabular}%
}
\caption{Comparison between scores of the 2D networks used in CHAOS challenge, the metrics values are drawn using the CHAOS online evaluation platform. The given values represent the prediction scores obtained using our method on each task along with the average scores on all tasks. The best results are given in bold. ASSD : Average symmetric surface distance; MSSD : Maximum symmetric surface distance; RAVD : Relative absolute volume difference.}
\label{tab:tab2}
\end{table*}

\begin{figure}[!h]
		\begin{minipage}[h]{\hsize}
		    \centering
			\includegraphics[height=\mysize cm, width=\mysize cm]{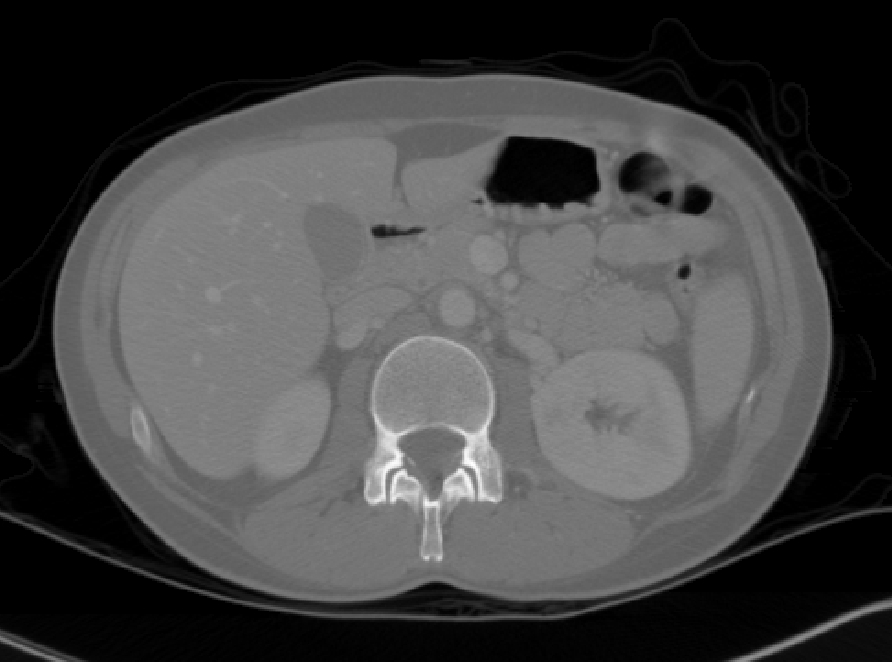}
			\includegraphics[height=\mysize cm, width=\mysize cm]{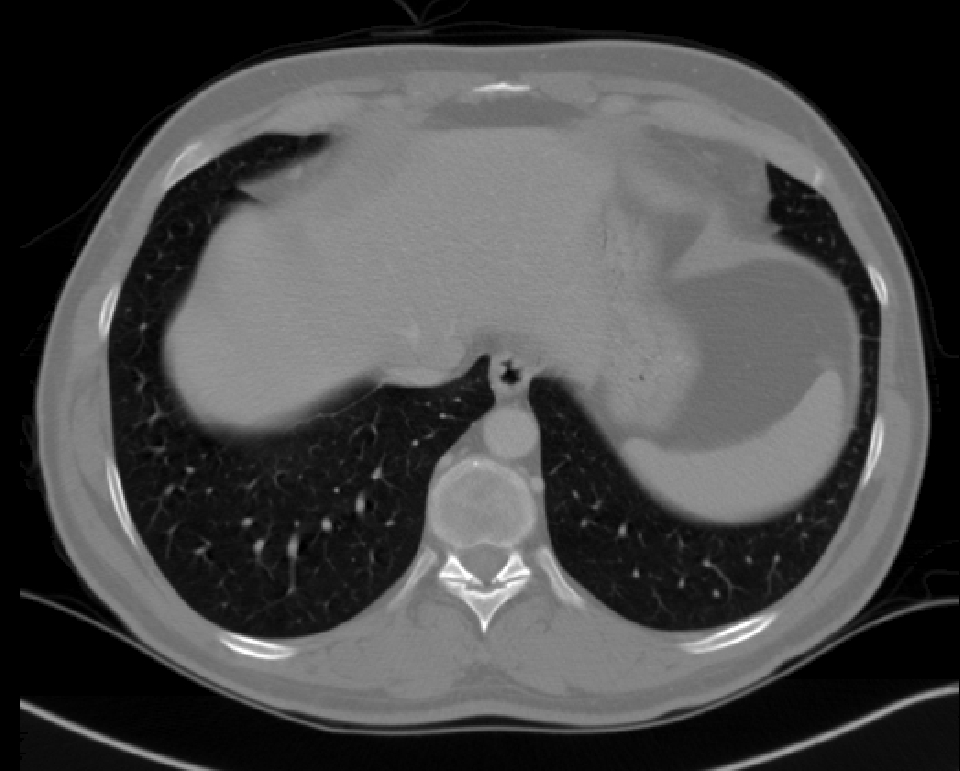}
			\vspace{1mm}
		\end{minipage}
		\begin{minipage}[]{\hsize}
			\centering
			\includegraphics[height=\mysize cm, width=\mysize cm]{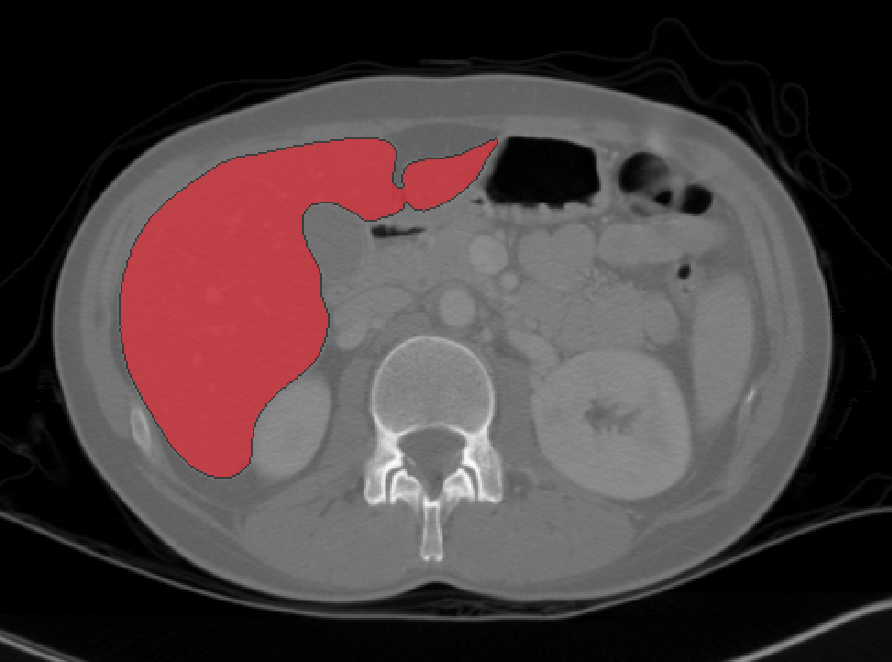}
			\includegraphics[height=\mysize cm, width=\mysize cm]{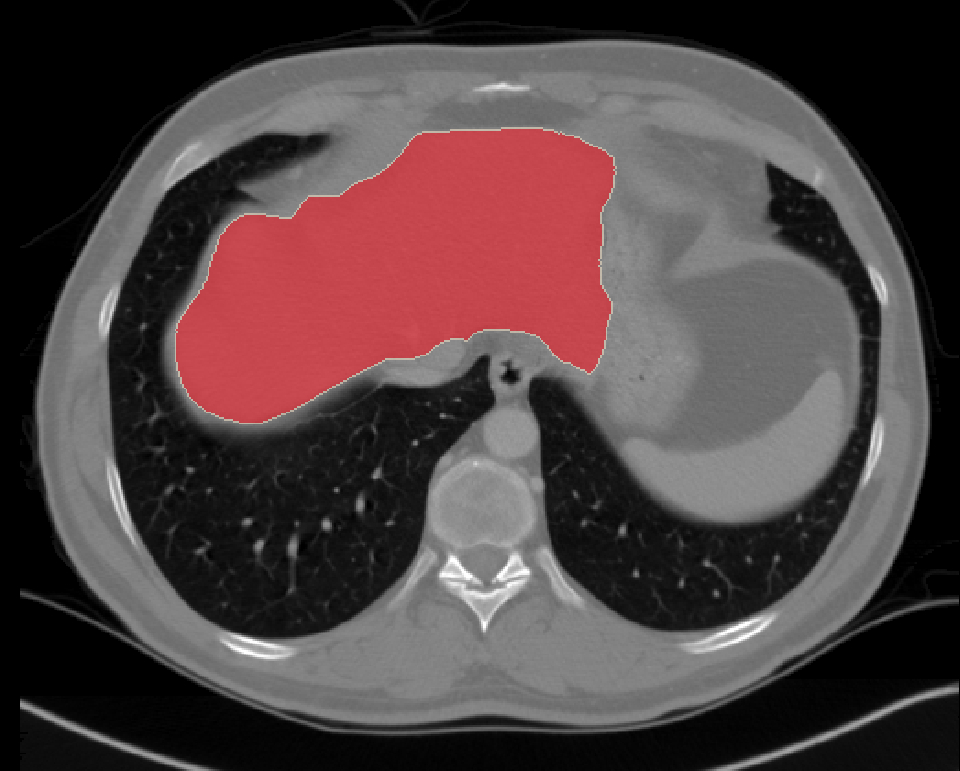}
			\vspace{1mm}
		\end{minipage}
	\caption{Visualizations on two samples from the CHAOS test set. Top to bottom: CT test data and their corresponding mask predicted by Omnia-Net.}
	\label{fig:chaosCT}
\end{figure}

Our approach is based on relatively the same architecture as described in Sect.~\ref{sssec:num1}, the only modification made is the substitution of the encoder by the B4 version. We employ the Nadam optimizer with an initial learning rate of 10$^{-3}$ to avoid an auto-restriction of the evolution of the model optimization to a local minimum. We opt for a learning rate decay method based on the reduction of the learning rate by 5\% at each epoch until reaching a minimal learning rate which in this case is at a level of 10$^{-5}$. This method aims at accelerating the convergence of the network in the first epochs of the training. The reduction of the learning rate to a critical minimum value mitigates the appearance of oscillations at later stages of the training which can occur when using a constant and relatively high learning rate.

In our experiments, we use a single network throughout the challenge tasks without using additional data during training. There are two different modalities in the CHAOS dataset, CT and MR (T1-DUAL and T2-SPIR sequences). We use a unique 2D network for the 5 different tasks and compare our results to the 2D networks that participated in the 5 tasks which are depicted in the challenge paper. As shown in Tab.~\ref{tab:tab2}, our network obtains the best averaged scores and particularly stands out in the multi-modal setting of Task 1, where we can see that our network shows a remarkable generalization performance with a gap of more than 11\% on the Dice score metric and more than 17 on the average metric, and this when compared to the best scores of the 2D networks shown in the CHAOS challenge paper. This can be due to several factors: we used larger image sizes (512$\times$512) contrary to the OvGUMEMoRIAL (128x128), IITKGP-KLIV (256$\times$256) and PKDIA (256$\times$256 MR, 512$\times$512 CT) teams which allowed a more stable training and better use of the learning potential of the network through the use of a constant input image resolution. It is also worth mentioning that we used a batch size of 8 unlike the others and the Nadam optimizer which performs better than the traditional Adam for computer vision tasks. 

\begin{figure}[!h]
    	\begin{minipage}[]{\hsize}
    	    \centering
    		\includegraphics[height=\mysize cm, width=\mysize cm]{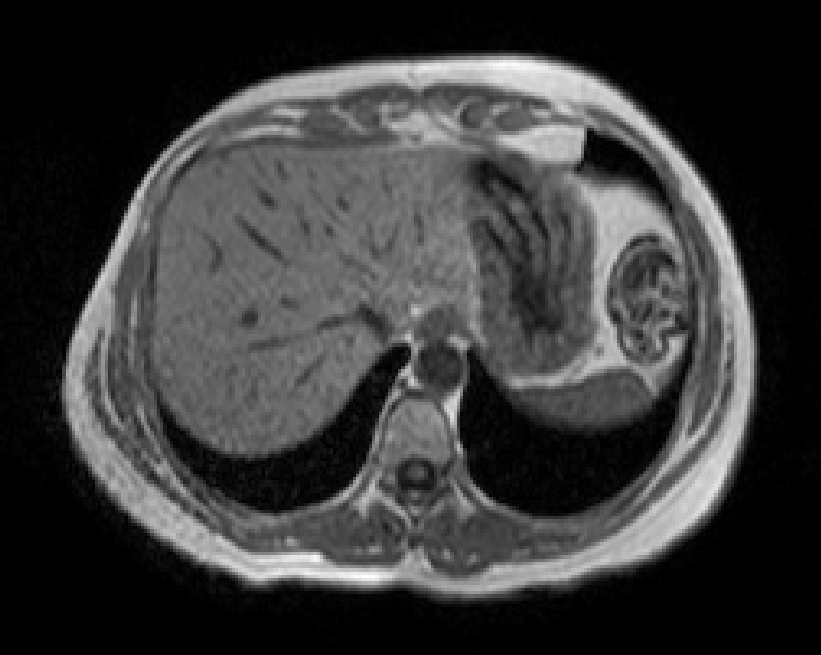}
    		\includegraphics[height=\mysize cm, width=\mysize cm]{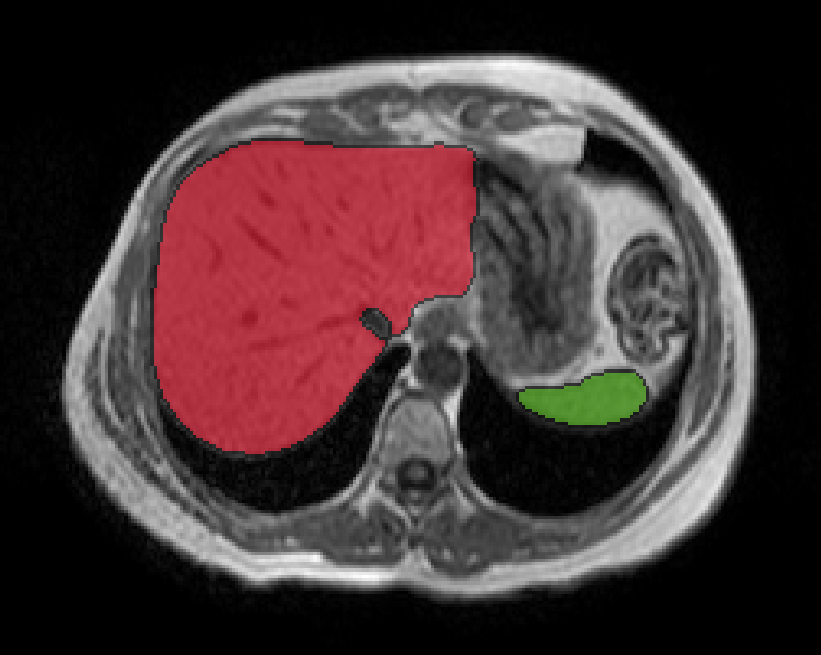}
    		\vspace{1mm}
    	\end{minipage}
    	\begin{minipage}[]{\hsize}
    	    \centering
    		\includegraphics[height=\mysize cm, width=\mysize cm]{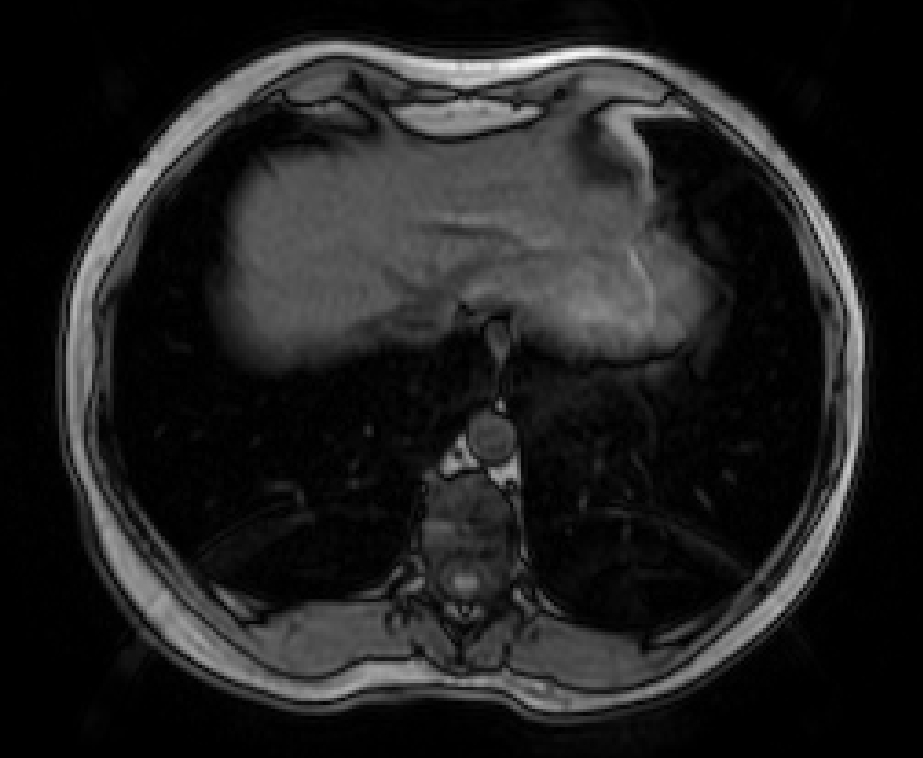}
    		\includegraphics[height=\mysize cm, width=\mysize cm]{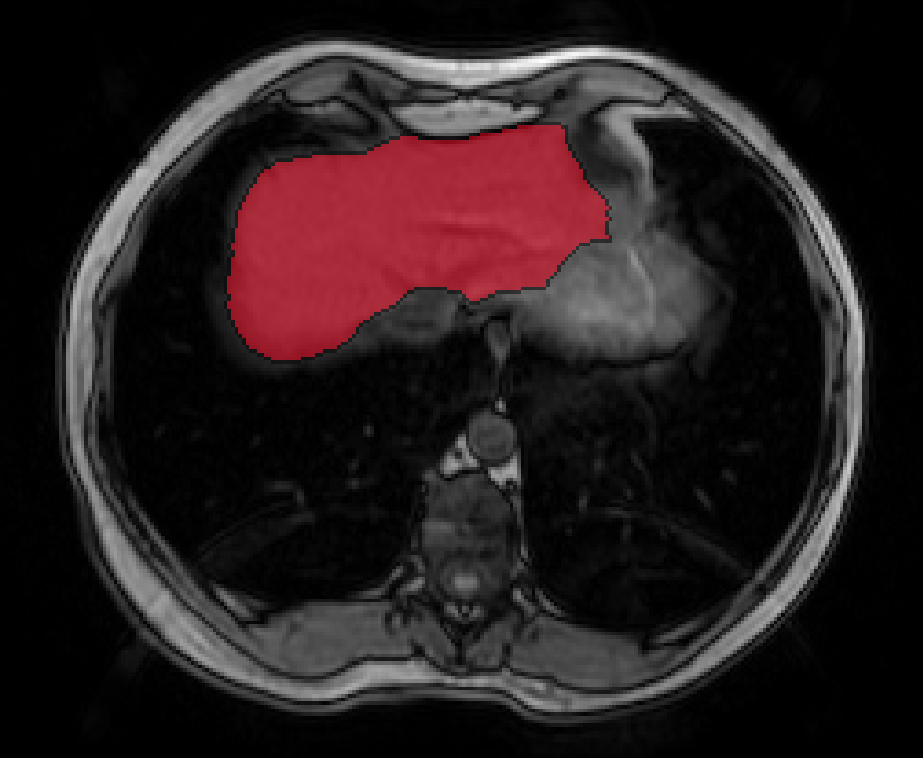}
    		\vspace{1mm}
    	\end{minipage}
    	\begin{minipage}[]{\hsize}
    	    \centering
    		\includegraphics[height=\mysize cm, width=\mysize cm]{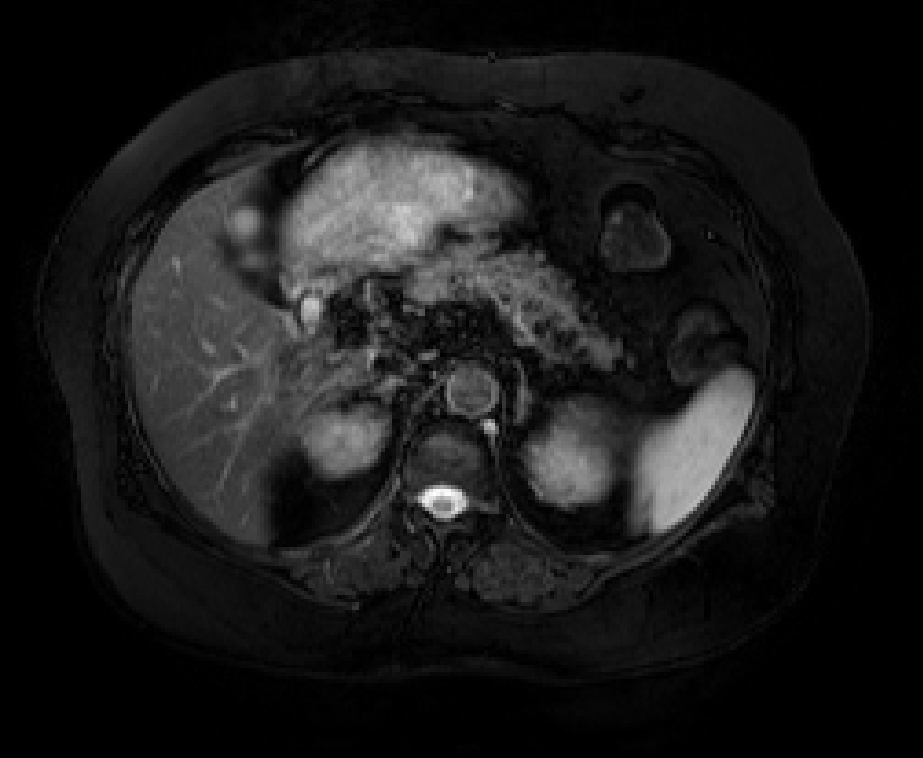}
    		\includegraphics[height=\mysize cm, width=\mysize cm]{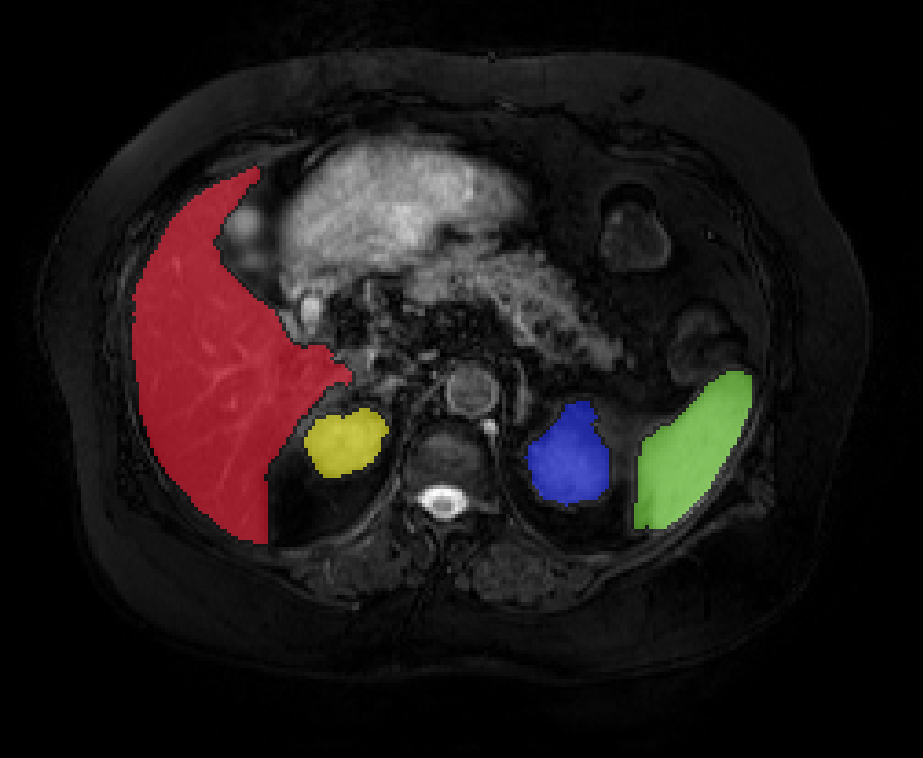}
    		\vspace{1mm}
    	\end{minipage}
	\caption{Visualizations on three samples from the CHAOS test set (Task 5). From top to bottom: liver, right and left kidney, and spleen in T1-in, T1-out, and T2-SPIR along with their according delineation prediction by Omnia-Net, are shown respectively in red, yellow, blue, and green.}
	\label{fig:Chaos2}
\end{figure}

The network we used is a key element of the performance of our method. We employed the EfficientNet as an encoder whose transfer power is no longer to be proven, we nevertheless added a convolution layer with 16 filters at the beginning of the encoder aiming at improving the performance of full-scale feature extraction. The output of this layer is then sent through skip-connection to the end of the decoder as shown in Fig.~\ref{fig:omnianet} to allow a better characterization of the segmented elements and to avoid any loss of information or noise that could be generated by the processing of data through the network. This was mainly motivated by the fact that the first layer of the EfficientNet directly compresses the input signal and does not take advantage of the full-scale information. Tab.~\ref{tab:tab2} also illustrates the network's ability to achieve a high performance that is generalizable to binary and multi-class cases, in both single and multiple modalities, which favors its adoption for real-life systems.

\subsubsection{3D multi-modal brain tumor MR segmentation}

Our segmentation approach follows an encoder-decoder CNN-based architecture with an embedded pre-trained encoder as introduced in~\citep{Hmessaoudi2021}. We perform a standard pre-processing strategy and crop the image in order to retain only the voxels containing the brain region and use a patch size of 128$\times$128$\times$128 for training. We then normalize with z-score the non-zero voxels of each input data channel independently and use a batch size of 1 to fit GPU memory constraints.

\begin{table}[!h]
\centering

    \begin{tabular}{llll}
    \hline
    Dice Score  & WT    & TC    & ET    \\ \hline
    Mean        & 91.69 & 83.23 & 81.75 \\ \hline
    Median      & 93.45 & 93.10 & 89.17 \\ \hline
    25 Quartile & 89.51 & 83.89 & 81.98 \\ \hline
    75 Quartile & 96.09 & 96.34 & 94.11 \\ \hline
    \end{tabular}%

\caption{DS-Net segmentation scores on the BraTS 2022 validation set.}
\label{tab:tab3}
\end{table}

Instead of minimizing the loss function using the errors obtained from the NET/Ncr, ED, and ET classes, we follow a region-based strategy in which the regions of interest become TC, WT, and ET. These regions are the ones that are scored by the challenge evaluation platform. In doing so, we use a Sigmoid activation function instead of Softmax at the end of the network since the new target classes are no longer independent. We train our network for 90 epochs and binarize the predicted outputs using a threshold of 0.5 and as for~\cite{carre21}, we keep the ET unaltered and extract the NET/Ncr from the TC and ED from the WT through logical operations. 

For network training, we use the sum of Dice and binary cross-entropy loss as a loss function, which operates on three class labels: TC, WT, and ET. CNNs often show a relative improvement during the training process by performing an alteration of the learning rate. Inspired by the step decay strategy of~\cite{he_steps} and the cosine annealing strategy of~\cite{Cos_Loshchilov}, we train our network with a constant learning rate of $10^{-3}$ without calculating the validation score until training score becomes greater than $0.85$. After 40 epochs, we use the cosine annealing strategy for the rest of the training with a minimum learning rate of $10^{-5}$. We opt for the use of the Nadam optimizer with coefficients betas set to (0.95, 0.99) along with LookAhead \citep{lookahead}. The number of fast weights updates is set to 6, and the magnitude of the final parameters (also called LookAhead parameter) is set to 0.5. 

Taking into account that many LGG data show an absence of enhancing tumor class, thus resulting in a high probability of a false positive that can be generated from the network during inference, we adopt a thresholding strategy targeting the ET class. If the ET volume is lower than a defined threshold, we ensure that the corresponding voxels are belonging to the NET/Ncr class in order for them to be calculated as part of the whole and core tumor. 

The results shown in Tab.~\ref{tab:tab3} underpin the performance of our method on the validation set. All segmentation results are evaluated on the BraTS 2022 challenge platform. The relatively low score of enhancing tumor when compared to other classes is mainly due to a binary score calculation system of the BraTS evaluation platform that penalizes the whole predicted class with the worst score if the latter includes a false positive voxel.

\begin{figure}[!h]
	\begin{minipage}[h]{\hsize}
	    \centering
		\includegraphics[height=\mysize cm, width=\mysize cm]{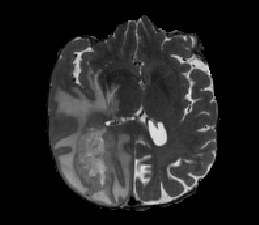}
		\includegraphics[height=\mysize cm, width=\mysize cm]{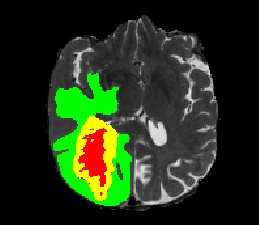}
		\vspace{1mm}
	\end{minipage}
	\begin{minipage}[]{\hsize}
	    \centering
		\includegraphics[height=\mysize cm, width=\mysize cm]{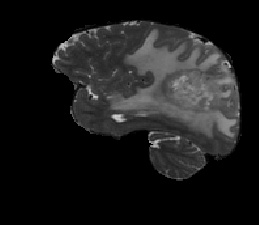}
		\includegraphics[height=\mysize cm, width=\mysize cm]{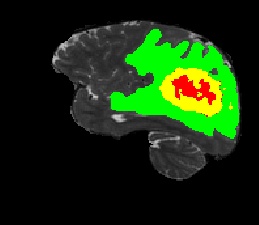}
		\vspace{1mm}
	\end{minipage}
	\begin{minipage}[]{\hsize}
		\centering
		\includegraphics[height=\mysize cm, width=\mysize cm]{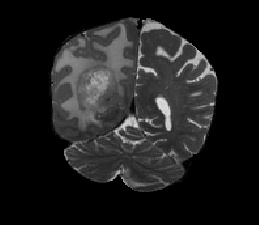}
		\includegraphics[height=\mysize cm, width=\mysize cm]{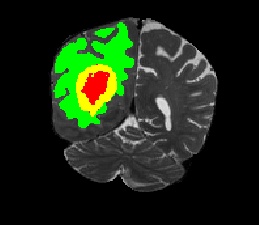}
		\vspace{1mm}
	\end{minipage}
	\caption{Visualizations on the $00015$ patient T2 data from the BraTS 2022 validation set. From top to bottom: Axial, sagittal and coronal views are shown with their corresponding DS-Net segmentation result. Edema is shown in green, enhancing tumor in yellow, and necrosis/non-enhancing tumor in red.} 
	\label{fig:Bratsvalid}
\end{figure}

The scores are relatively better for other classes. Our network more easily distinguishes peritumoral edema and NET/Ncr volumes as can be seen in Fig.~\ref{fig:Bratsvalid} which depicts the segmentation results over some cases from the validation set. It can be seen that our network can generate convincing results, even on some small volumes which can be due to the exploitation of the second dimension leading to a better feature extraction of the intra-slice information and thus showing the effectiveness of our approach on brain tumor segmentation problems.

\begin{figure}[p]
  \begin{minipage}[]{\hsize}
		\centering
		\textbf{Best :}  BraTS22 Validation 00190,  TC = 0.9930, WT = 0.9883, ET = 0.9748\\
		\vspace{1mm}
		\includegraphics[height=\mysize cm, width=\mysize cm]{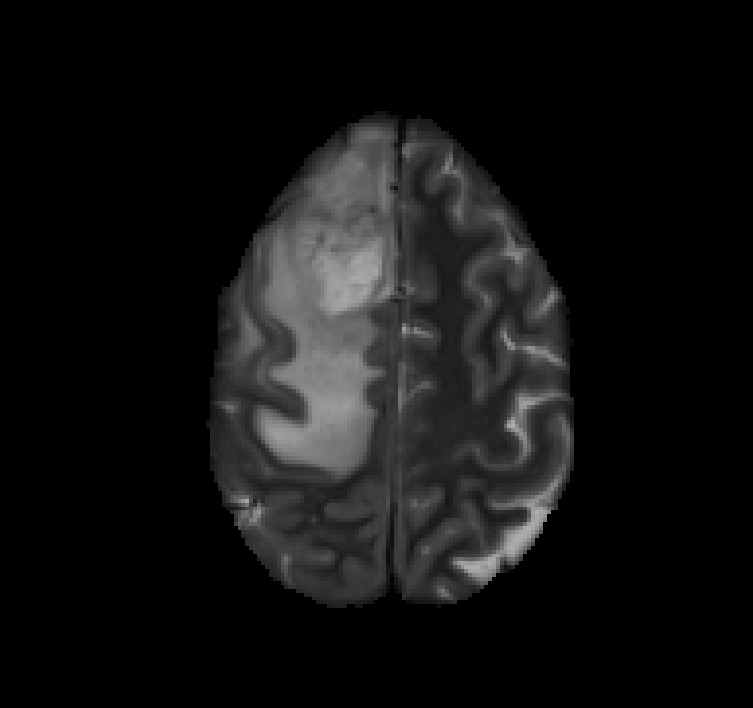}
		\includegraphics[height=\mysize cm, width=\mysize cm]{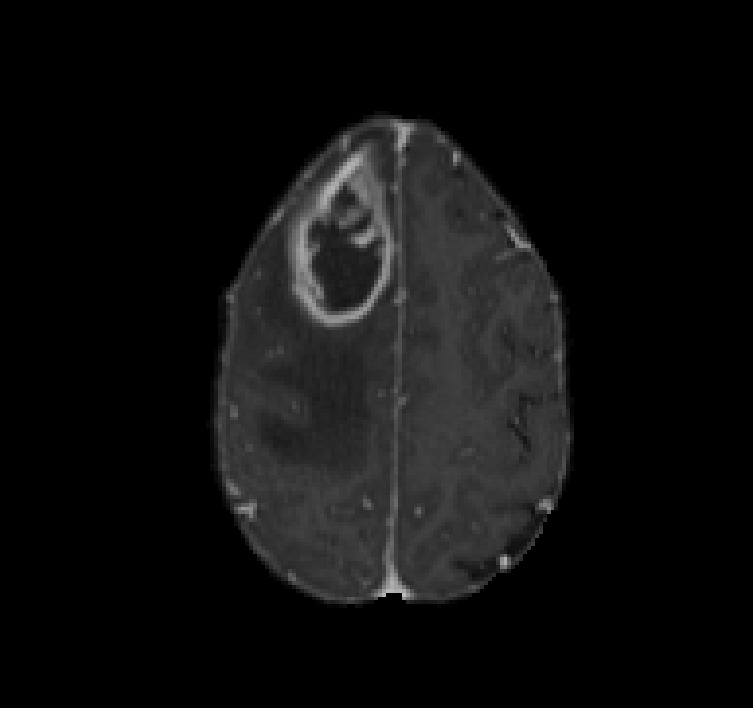}
		\includegraphics[height=\mysize cm, width=\mysize cm]{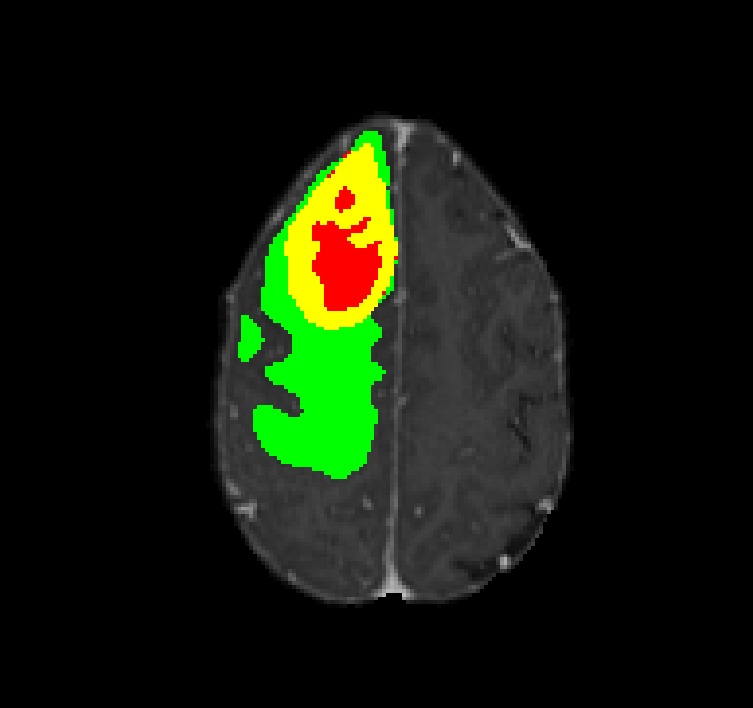}
		\vspace{1mm}
	\end{minipage}
	\begin{minipage}[]{\hsize}
		\centering
		\textbf{25th Percentile :}   BraTS22 Validation 01719,  TC = 0.7868, WT = 0.7927, ET = 1.0\\
		\vspace{1mm}
		\includegraphics[height=\mysize cm, width=\mysize cm]{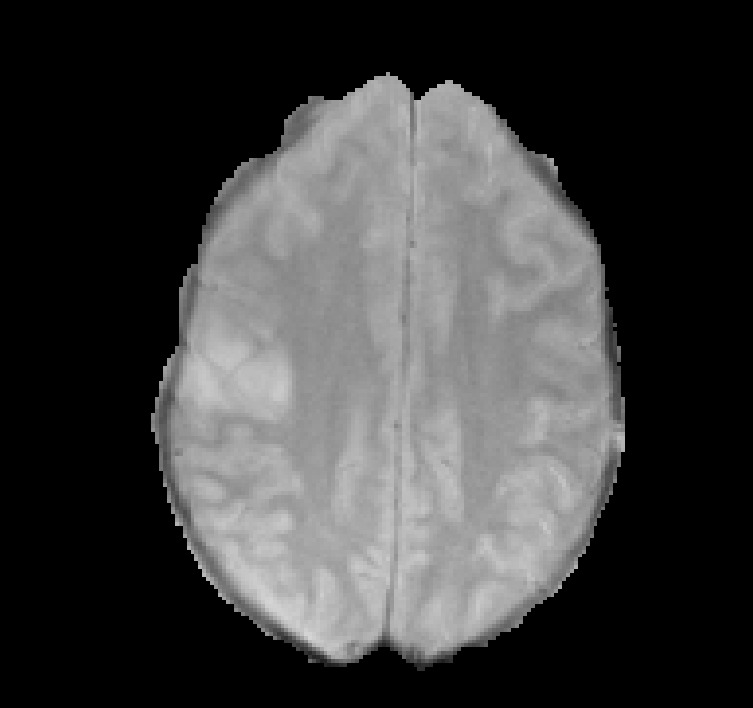}
		\includegraphics[height=\mysize cm, width=\mysize cm]{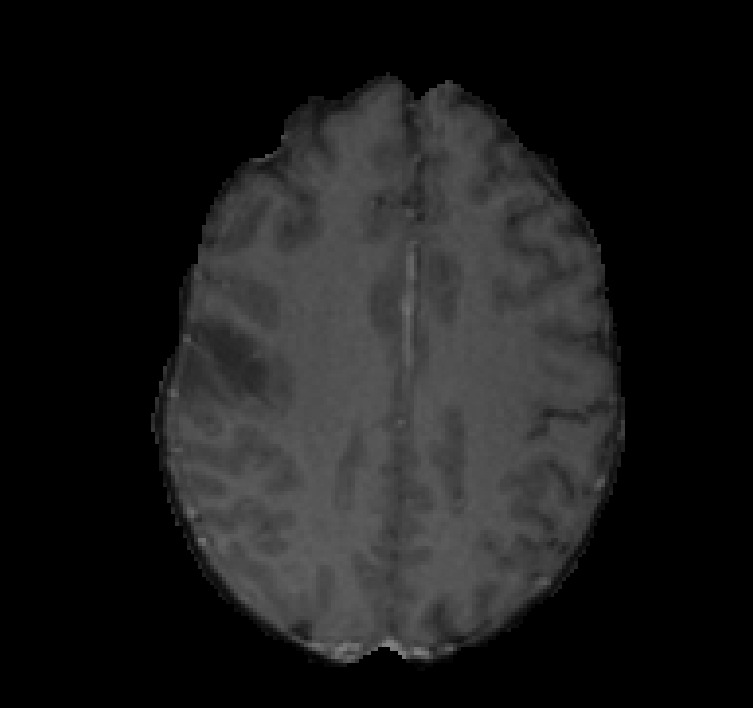}
		\includegraphics[height=\mysize cm, width=\mysize cm]{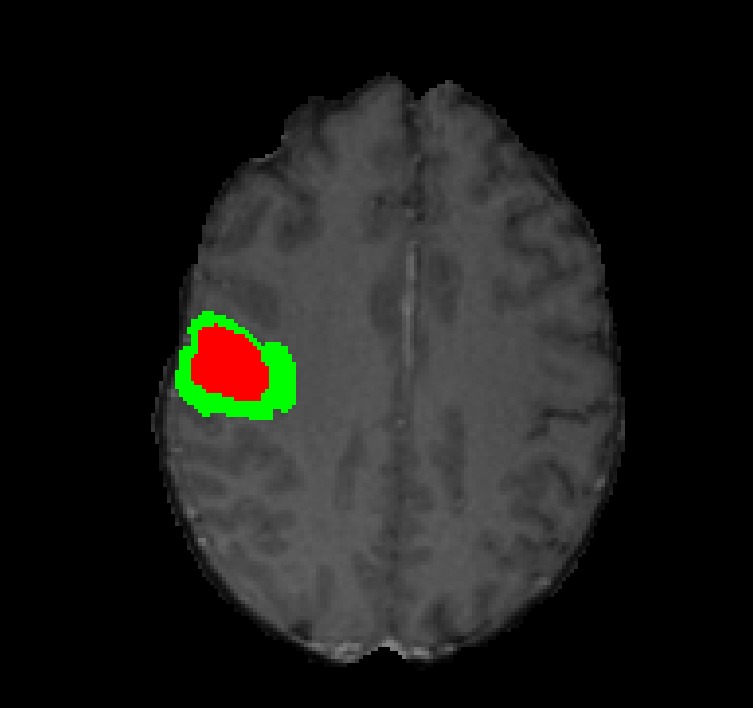}
		\vspace{1mm}
	\end{minipage}
	\begin{minipage}[]{\hsize}
		\centering
		\textbf{Median  :}   BraTS22 Validation 00462,  TC = 0.9569, WT = 0.9047, ET = 0.8865\\
		\vspace{1mm}
		\includegraphics[height=\mysize cm, width=\mysize cm]{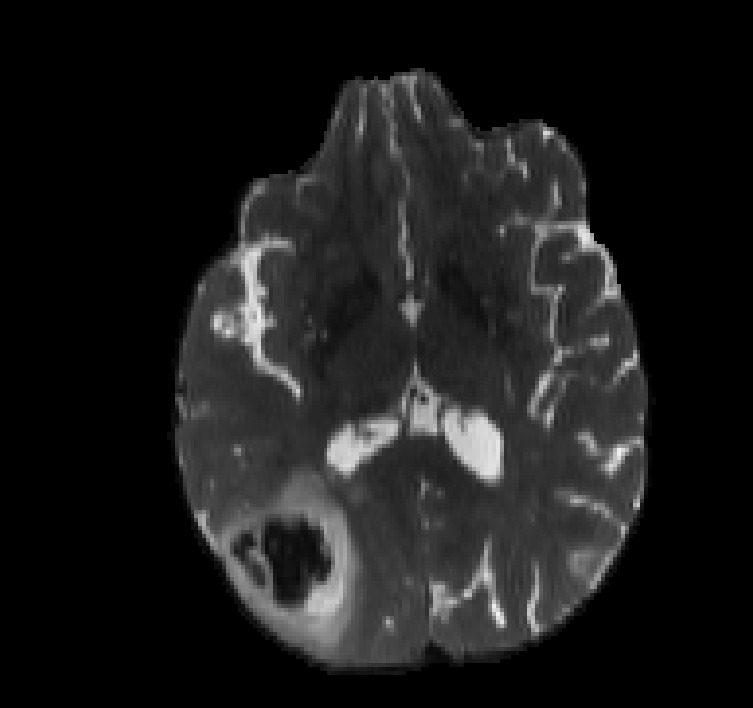}
		\includegraphics[height=\mysize cm, width=\mysize cm]{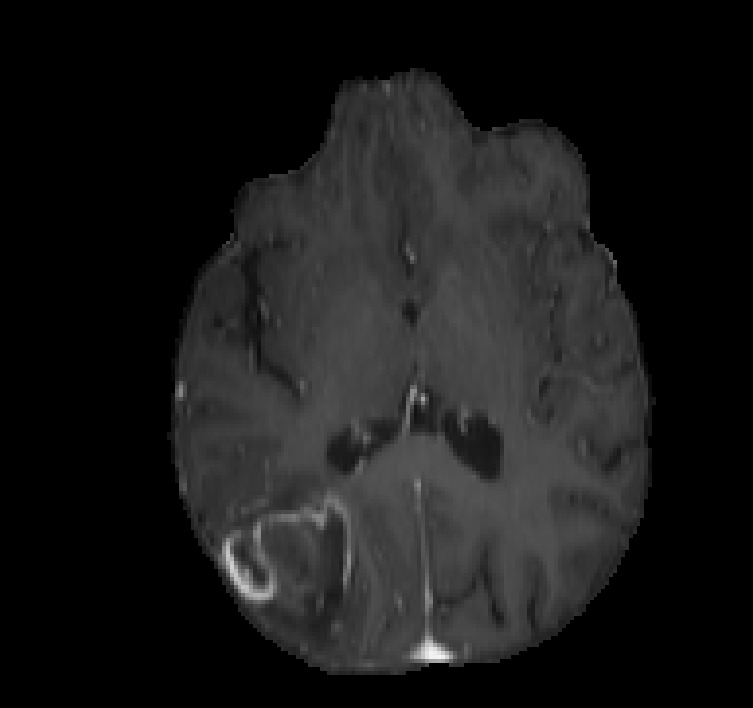}
		\includegraphics[height=\mysize cm, width=\mysize cm]{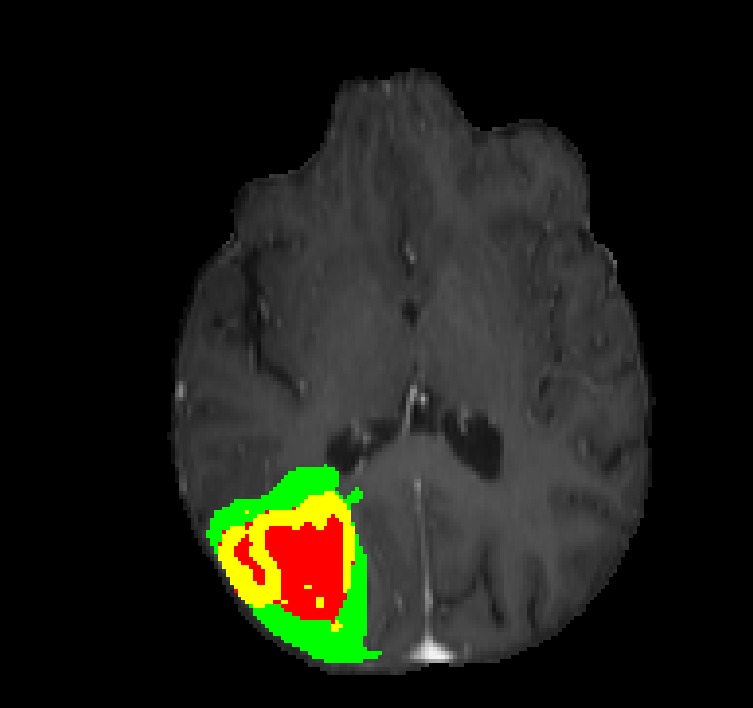}
		\vspace{1mm}
	\end{minipage}
	\begin{minipage}[]{\hsize}
		\centering
		\textbf{75th Percentile :}   BraTS22 Validation 00553,  TC = 0.9709, WT = 0.9452, ET = 0.9324\\
		\vspace{1mm}
		\includegraphics[height=\mysize cm, width=\mysize cm]{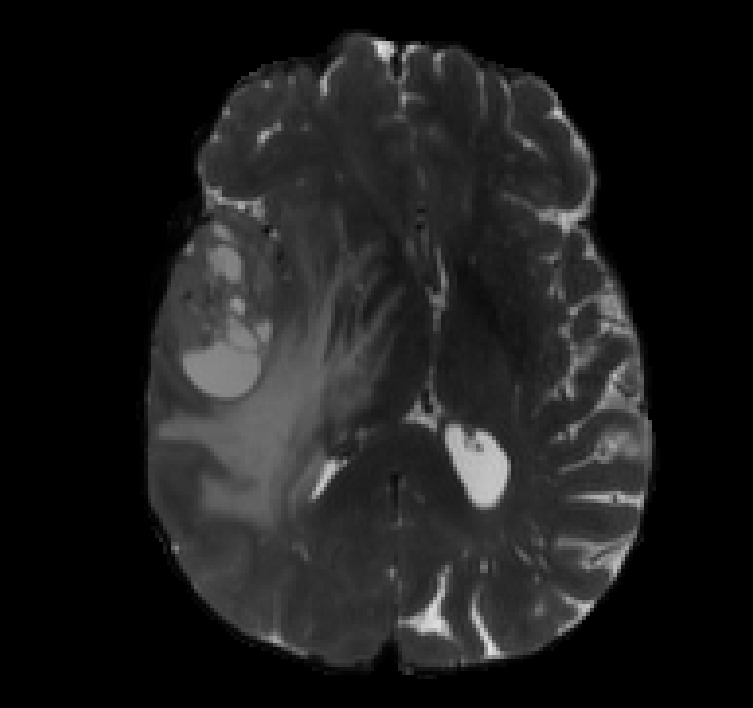}
		\includegraphics[height=\mysize cm, width=\mysize cm]{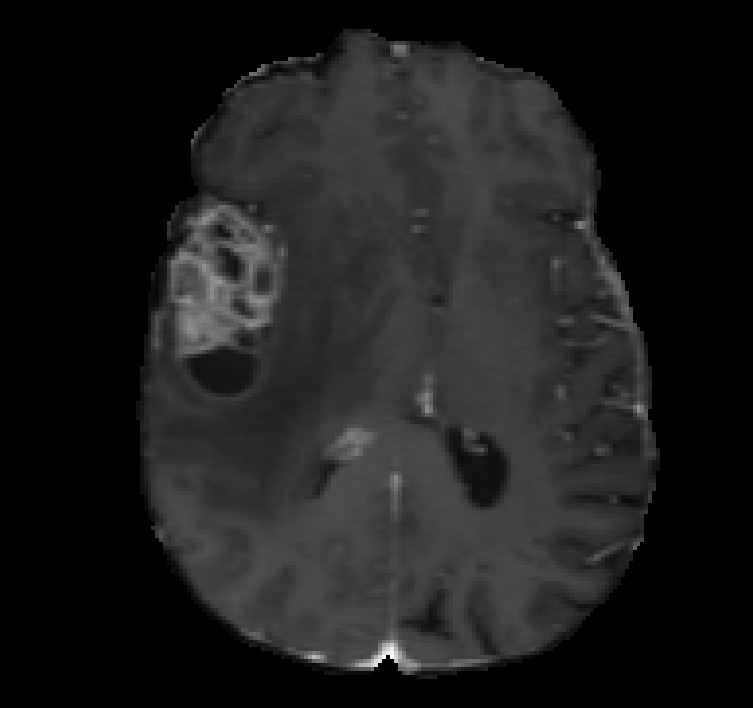}
		\includegraphics[height=\mysize cm, width=\mysize cm]{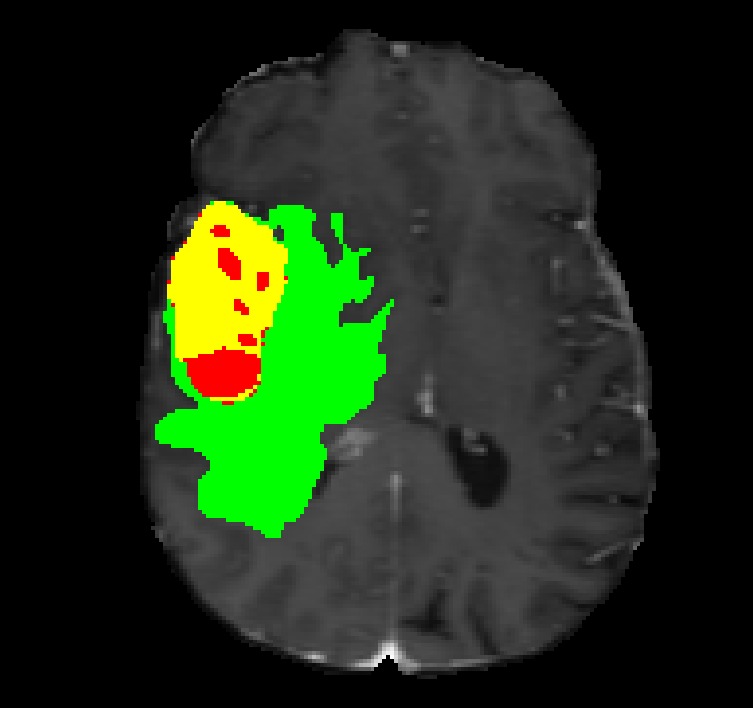}
		\vspace{1mm}
	\end{minipage}
	\begin{minipage}[]{\hsize}
		\centering
		\textbf{Worst :}   BraTS22 Validation 00213,  TC = 0.0 , WT = 0.2394 , ET = 0.0\\
		\vspace{1mm}
		\includegraphics[height=\mysize cm, width=\mysize cm]{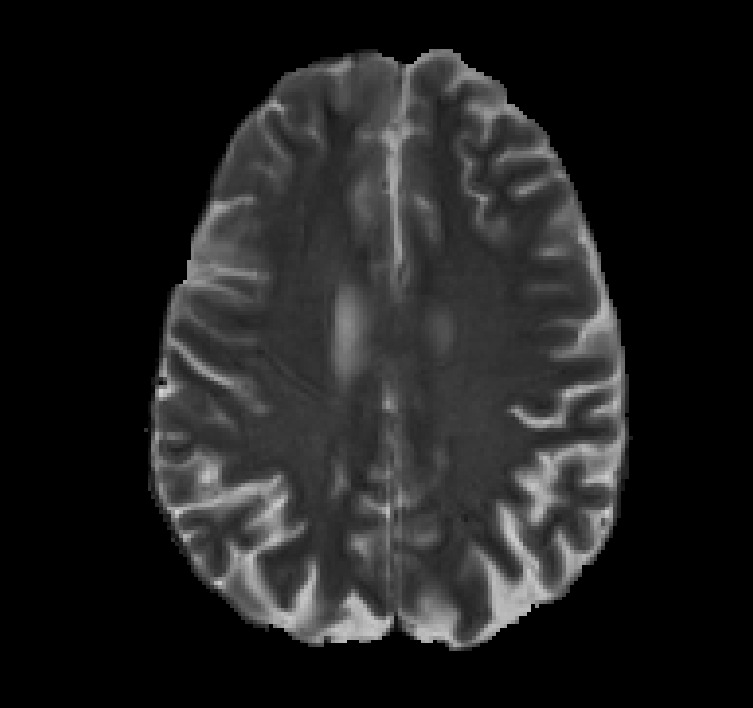}
		\includegraphics[height=\mysize cm, width=\mysize cm]{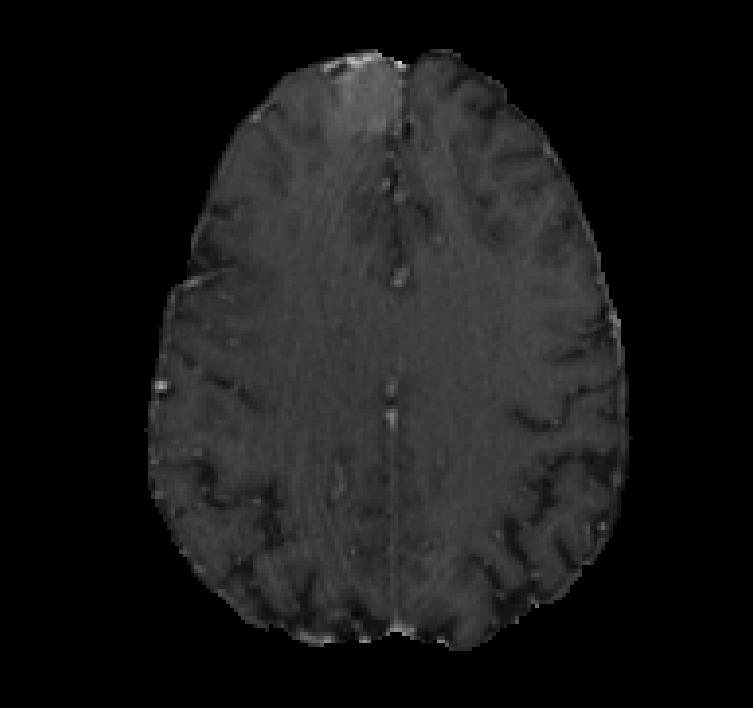}
		\includegraphics[height=\mysize cm, width=\mysize cm]{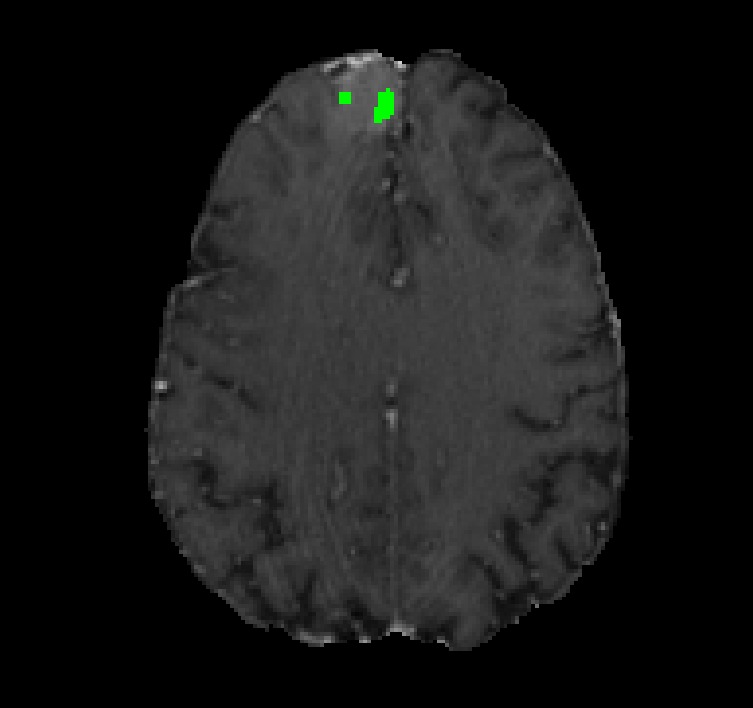}
		\vspace{1mm}
	\end{minipage}
	\caption{DX-Net qualitative results. Cases were selected as best, worst, median
and 75th and 25th percentile. Within each row, the raw T2 image is shown to the left, the T1ce image in the middle and on overlay with the generated segmentation on the T1ce image is shown on the right. Edema is shown in green, enhancing tumor in yellow and necrosis/non-enhancing tumor on red.}
	\label{fig:bratsminmax}
\end{figure}

\subsection{Dimensional transfer learning}

\subsubsection{3D multi-modal brain tumor MR segmentation}

The two networks, DS-Net and DX-Net, are both designed for 3D image segmentation tasks. They differ in their encoding and decoding processes, with DX-Net having a 3D pre-trained encoder and a 3D decoder, and DS-Net having a 3D encoder followed by a 2D pre-trained encoder, a 2D decoder, and a 3D decoder. The training process undertaken for the DX-Net is similar to the one followed for DS-Net. From Tab.~\ref{tab:tab3}, we can see the mean Dice score of each region computed by the online evaluation platform. From it, we can derive that the whole tumor region is more easily detected by both networks. However, if we look at the TC and ET scores for both networks, we can see that the peritumoral edema is much better segmented by DS-Net, this can be deduced by looking at the superiority of the mean Dice of ET and TC for DX-Net when compared to that of the DS-Net, despite that DS-Net shows a higher score for the whole tumor region which indicates a much better delineation of the peritumoral edema class produced by DS-Net. Nevertheless, the DX-Net shows a clear superiority in the discrimination of the tumor core and enhancing tumor region. We assume that the underlying cause could be a better exploitation of volumetric features, and this is through the application of a deeper and better initialized 3D encoder. DX-Net post-processed predictions are shown in Fig.~\ref{fig:bratsminmax}. 

The use of a fully 3D network, as in DX-Net, allows for the capture of authentic volumetric features in the image, which can lead to improved segmentation, especially for difficult areas in the image. This is because 3D convolution layers can learn the inter-slice relationships within the depth dimension using the generated pre-trained 3D weights, allowing for a more reliable representation of the input image. On the other hand, DS-Net's approach of first encoding the 3D data into 2D data and then processing it with a 2D pre-trained encoder, may be slower and sub-optimal for configurations requiring fast training as compared to DX-Net. This is evident from the training time, where DS-Net takes 2 seconds per image to train, while DX-Net takes 1.5 seconds per image on an NVIDIA Tesla P100 GPU. However, this trade-off in computational efficiency may come at the cost of not fully capturing the intra-slice information as efficiently as by DS-Net.

\begin{table}[!h]
\centering

    \begin{tabular}{llll}
    \hline
    Dice Score  & WT    & TC    & ET    \\ \hline
    Mean        & 91.22 & 84.77 & 83.88 \\ \hline
    Median      & 94.15 & 93.46 & 89.74 \\ \hline
    25 Quartile & 90.08 & 85.56 & 83.49 \\ \hline
    75 Quartile & 96.46 & 96.48 & 94.92 \\ \hline
    \end{tabular}%

\caption{DX-Net segmentation scores on the BraTS 2022 validation set.}
\label{tab:tab4}
\end{table}

Furthermore, the networks could be significantly improved by extending the training time and also benefiting from a more aggressive data augmentation scheme. The learning process also depends mostly on the annotations performed, these last ones can undermine the proper generalizability of the network, especially when the task is complex and error-prone. The networks are trained for less than 100 epochs and without extensive post-processing. The particular strength of the DX-Net is its relatively low computational power consumption and its ability to accurately retain and reproduce expert annotations, which might motivate its use in a clinical context.


\section{Conclusion}
\label{sec:conclusion}

In this work, we introduced two transfer learning paradigms. First, we presented the weight transfer learning, an efficient approach to re-use the weights of a pre-trained 2D classifier network by embedding it in a network of the same or higher dimension. We derived from it two network architectures: Omnia-Net, a 2D network which is intended for the segmentation of 2D echo-cardiographic and 3D MR and CT data and DS-Net, a 3D network embedding a 2D architecture that allows the exploitation of intra-slice information from brain tumor images.

The second proposed approach is the dimensional transfer learning, which is based on the 3D weights extrapolation of a pre-trained 2D network. DX-Net is a network derived from this approach. It is a U-Net-like architecture using a 3D EfficientNet encoder. The latter is initialized with extrapolated 3D weights from the 2D EfficientNet weights, which are pre-trained on ImageNet following a noisy-student strategy.

Empirical results showed that our approaches clearly outperform state-of-the-art methods. Omnia-Net ranked first in the CAMUS challenge, which involved addressing variable image quality, multi-chamber views, and multi-phase context. This network produced convincing results and can already be used in a clinical context. On the CHAOS challenge, our approach was ranked 3$^{rd}$ with the same network and got better results than all methods using 2D networks
in the challenge. DS-Net and DX-Net showed promising results
and achieve competitive performance. Our future perspectives will be turned towards the investigation of other methods to handle weight transformation from 2D to higher dimensions in the context of dimensional transfer learning to achieve better performance conservation in higher dimensions and enhanced image segmentation quality.

\section{Declaration of interests}
The authors declare that they have no known competing financial interests or personal relationships that could have appeared to influence the work reported in this paper.

\section{Acknowledgment}
This work is sponsored by the General Directorate for Scientific Research and Technological Development, Ministry of Higher Education and Scientific Research (DGRSDT), Algeria. 

\section{Funding}
This research did not receive any specific grant from funding agencies in the public, commercial, or not-for-profit sectors.

\bibliographystyle{model2-names}
\bibliography{article}

\end{document}